\documentclass[pra,aps,10pt,twocolumn,superscriptaddress,showpacs,floatfix,longbibliography,citeautoscript]{revtex4-1}

\usepackage{graphicx}
\usepackage{color}
\usepackage{latexsym}
\usepackage{amsmath}
\usepackage{amsfonts}
\usepackage{amssymb}
\usepackage{subfigure}
\usepackage{bm}
\usepackage{float}

\usepackage{hyperref}
\hypersetup{colorlinks=true}
\usepackage[all]{hypcap} 

\newcommand{\varg}{\lambda}

\begin{document}

\title{
Enhanced nonlinear interaction effects in a four-mode optomechanical ring
}

\author{Li-Jing Jin} 
\affiliation{Beijing Computational Science Research Center, Beijing 100193, China}

\author{Jing Qiu} 
\affiliation{Beijing Computational Science Research Center, Beijing 100193, China}

\author{Stefano Chesi}
\email{stefano.chesi@csrc.ac.cn}
\affiliation{Beijing Computational Science Research Center, Beijing 100193, China}

\author{Ying-Dan Wang}
\email{yingdan.wang@itp.ac.cn}
\affiliation{CAS Key Laboratory of Theoretical Physics, Institute of Theoretical Physics, Chinese Academy of Sciences, P.O. Box 2735, Beijing 100190, China}
\affiliation{School of Physical Sciences, University of Chinese Academy of Sciences, No.19A Yuquan Road, Beijing 100049, China}
\affiliation{Synergetic Innovation Center for Quantum Effects and Applications, Hunan Normal University, Changsha 410081, China}

\begin{abstract}
With a perturbative treatment based on the Keldysh Green's function technique, we study the resonant enhancement of nonlinear interaction effects in a four-mode optomechanical ring. In such a system, we identify five distinct types of resonant scattering between unperturbed polariton modes, induced by the nonlinear optomechanical interaction. By computing the cavity density of states and optomechanical induced transparency signal, we find that the largest nonlinear effects are induced by a decay process involving the two phonon-like polaritons. In contrast to the conventional two-mode optomechanical system, our proposed system can exhibit prominent nonlinear features even in the regime when the single-photon coupling is much smaller than the cavity damping.
\end{abstract}

\date{\today}

\maketitle

\section{Introduction}

Benefiting from the development of semiconductor fabrication and nanotechnology, cavity optomechanical systems are receiving increasing attention due to both fundamental reasons and potential applications \cite{aspelmeyer2014cavity,aspelmeyer2010quantum,bowen2015quantum,kippenberg2007cavity,meystre2013short}.  In particular, they provide a promising platform not only for exploring macroscopic quantum mechanical behaviors \cite{marshall2003towards,vitali2007optomechanical,romero2011large,pikovski2012probing}, but also for sensitive measurements \cite{rugar2004single} and quantum information processing \cite{mancini2003scheme,stannigel2010optomechanical}. 

In cavity optomechanical systems, the interactions between single photon modes and mechanical modes are intrinsically nonlinear \cite{aspelmeyer2014cavity}. Such nonlinear property is able to trigger various interesting phenomena, like non-classical states \cite{bose1997preparation}, photon blocking \cite{rabl2011photon}, cavity density of states (DOS) splitting \cite{lemonde2013nonlinear,borkje2013signatures}, and so on. So far, however, experiments have mostly investigated the linear regime, where significant achievements include the obsevation of normal-mode splitting \cite{groblacher2009observation}, cooling the mechanical mode to the ground state \cite{teufel2011sideband,chan2011laser}, optomechanically induced transparancy (OMIT) \cite{weis2010optomechanically,safavi2011electromagnetically}, as well as  the demonstration of squeezed light \cite{brooks2012non,safavi2013squeezed,purdy2013strong}.

Nonlinear features have been explored to a much lesser extent because they are typically suppressed by the ratio $g/\omega_m$ between the weak bare optomechanical coupling $g$ and the mechanical frequency $\omega_m$ \cite{rabl2011photon, nunnenkamp2011single}.  Using lasers to drive cavities, the coupling between photon modes and mechanical modes can be enhanced effectively, but one has to pay the price of  smeared nonlinear effects. Therefore, in order to observe nonlinear signatures, the coupling strength $g$ between a single photon and the mechanical motion should be sufficiently large. 
In the past years,  great progress towards larger $g$ had been obtained  in various optomechanical devices, e.g., ultracold
atoms in optical resonators \cite{gupta2007cavity, murch2008observation}, optomechanical crystals \cite{eichenfield2009optomechanical}, as well as superconducting circuits \cite{teufel2011circuit}. This provides a promising route for entering the nonlinear regime.

On the theoretical side, proposals for enhancing nonlinear effects are mostly based on two-mode optomechanical systems \cite{lemonde2013nonlinear,borkje2013signatures, liu2013parametric, lemonde2016enhanced,heikkila2014enhancing}. In particular, previous works have identified an interesting regime where the nonlinear interaction causes the higher-frequency polariton (a joint photonic-phononic excitation) to scatter resonantly into two lower-frequency polaritons \cite{lemonde2013nonlinear,borkje2013signatures,liu2013parametric}. Then, the nonlinear signatures are controlled by the ratio $g/\kappa$. In the resolved side-band regime, i.e., $\kappa \ll \omega_m$, the condition $g \sim \kappa$ is much easier to achieve than $g \sim \omega_m$. 
However, although the nonlinear effects are admittedly greatly enhanced, to meet the required condition is still a big challenge for most optomechanical platforms.


Following these motivations, we propose here a four-mode cavity optomechanical ring in which the nonlinear effects can be significantly more pronounced than in a two-mode system. In particular, we expect large nonlinear signatures with more routinely realizable experimental parameters. The nonlinear features are captured by changes in either the cavity DOS or the OMIT signal.  

Without nonlinear effects, the DOS exhibits four pronounced Lorenzian peaks, corresponding to the normal modes of the optomechanical ring. After including the nonlinear interactions, the response of the normal modes is modified and we find that nonlinear effects become visible whenever the system meets one of five special resonant conditions. Among them, a purely intra-branch scattering process between phonon-like excitations plays a dominant role. In order to obtain a clear nonlinear signature, we suggest to perform the OMIT measurements with a varying detuning or driving strength, which leads to a pronounced peak (nonlinear signature) in the OMIT reflection probability.

The paper is organized as follows. In Sec.~\ref{Sec: the system} we introduce the system Hamiltonian. The normal modes and polariton damping rates are calculated in the linear regime in Sec.~\ref{Sec: linear}. In Sec.~\ref{Sec: nonlinear} we include a weak nonlinear interaction and study its effect on the cavity DOS. 
In Sec.~\ref{Sec: results and discussions} we present our main results and in Sec.~\ref{sec_2mode} we provide an explicit comparison of the two-mode and four-mode systems. A specific realization of the four-mode optomechanical model is discussed in Sec.~\ref{sec: Experimental realization}. Finally, we conclude in Sec.~\ref{Sec: summary} and provide additional technical details in Appendices~\ref{appendix1} and \ref{appendix2}.

\section{System}\label{Sec: the system}

\begin{figure}
\centering
\includegraphics[width=0.6\columnwidth]{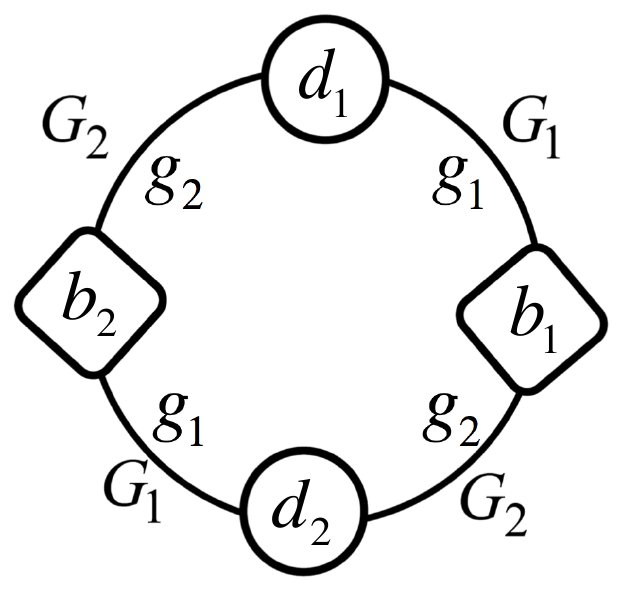}
\caption{Schematic illustration of the 4-mode optomechanical ring. It consists of two cavities (circles) and two mechanical oscillators (diamonds). $g_{1,2}$ and $G_{1,2}$ indicate the bare and dressed optomechanical couplings, respectively. 
}
\label{fig: the system}
\end{figure}

The system under consideration is illustrated in Fig.~\ref{fig: the system}: it is a periodic ring formed by two elementary optomechanical unit cells. While such configuration could be realized by several alternative implementations, a simple setup with two mechanical elements in a single cavity is explicitly discussed in Sec.~\ref{sec: Experimental realization}. 

Within each unit cell of the ring, the internal bare coupling between the photon mode of the cavity (frequency $\omega_c$, damping rate $\kappa$) and the phonon mode of the mechanical oscillator (frequency $\omega_m$, damping rate $\gamma$)  is $g_1$, while the bare optomechanical couplings between different cells are $g_2$. Driving the two cavities with identical lasers (frequencies $\omega_{\rm L}$), the effective couplings between photon modes and mechanical modes are enhanced to $G_1$ and $G_2$, respectively. Accordingly, the system Hamiltonian can be written into three parts \cite{aspelmeyer2014cavity}: 
\begin{equation}\label{H}
\hat{H}=\hat{H}_{\rm 0} + \hat{H}_{\rm nl} +\hat{H}_{\rm diss} ,
\end{equation} 
where the first term contains the linear part of the interaction. In a rotating frame at the laser frequency $\omega_{\rm L}$, and after a standard displacement transformation of the photon modes, $\hat{H}_0$ is given by ($\hbar =1$):
\begin{eqnarray} \label{Hamiltonian: linear}
\hat{H}_{\rm 0} &=& \sum_{n=1,2}\Big( \omega_m \hat{b}^{\dagger}_n \hat{b}_n -\Delta \hat{d}^{\dagger}_n \hat{d}_n + G_1 (\hat{d}_n + \hat{d}_n^{\dagger}) (\hat{b}_n + \hat{b}_n^{\dagger})\nonumber\\
&&+ G_2 (\hat{d}_n + \hat{d}_n^{\dagger}) (\hat{b}_{n-1} + \hat{b}_{n-1}^{\dagger})\Big),
\end{eqnarray}
where $\Delta = \omega_{\rm L} - \omega_{c}$ is the detuning of the two drives, $\hat{d}_{1,2}$ are the displacement operators describing the photon modes, and $\hat{b}_{1,2}$ represent the phonon modes. In Eq.~(\ref{Hamiltonian: linear}) we use the convention $\hat{b}_{0} \equiv \hat{b}_{2}$. The nonlinear term is:
\begin{equation}
\hat{H}_{\rm nl} = \sum_{n=1,2} \Big( g_1 \hat{d}_n^{\dagger} \hat{d}_n (\hat{b}_n +\hat{b}_n^{\dagger})+ g_2 \hat{d}_n^{\dagger} \hat{d}_n (\hat{b}_{n-1} + \hat{b}_{n-1}^{\dagger})\Big). \label{Hamiltonian: nonlinear}
\end{equation}
It should be noted that the couplings $G_{1,2}$ and $g_{1,2}$ are not independent, since:
\begin{equation}\label{ratio_Gg}
G_1/g_1 = G_2/g_2 =\sqrt{N_c},
\end{equation}
where $N_c \gg 1$ is the average number of photons in the driven cavities. 

Finally, $\hat{H}_{\rm diss}$ describes the effect of dissipative baths, which we assume Markovian over the frequencies of interest, and is given by: 
\begin{align} \label{H_diss}
& \hat{H}_{\rm diss}= \sum_{n,j} \left( \Delta_{n,j} \hat{o}^{\dagger}_{n,j} \hat{o}_{n,j} + i \sqrt{\frac{\kappa}{2 \pi \rho_c}} ( \hat{o}^{\dagger}_{n,j} \hat{d}_n - \hat{o}_{n,j} \hat{d}^{\dagger}_n ) \right)\nonumber \\
& +  \sum_{n,j} \left( \omega_{n,j} \hat{p}^{\dagger}_{n,j} \hat{p}_{n,j}  + i \sqrt{\frac{\gamma}{2 \pi \rho_m}}  (\hat{p}^{\dagger}_{n,j} - \hat{p}_{n,j} ) (\hat{b}_n + \hat{b}^{\dagger}_n)\right) ,
\end{align}
where $\hat{o}_{n,j}$ ($\hat{p}_{n,j}$) are the optical (phononic) bath operators for mode $n=1,2$. For simplicity, we take all the optical (mechanical) baths with identical linewidths $\kappa$~($\gamma$) and density of states $\rho_c$ ($\rho_m$). Like the cavity modes, the optical baths are written in a frame rotating at frequency $\omega_L$ (the original bath frequencies are $\Delta_{n,j}+\omega_L$). In the first line we have neglected the fast-oscillating interaction terms proportional to $e^{\pm 2i \omega_L t }$. Instead, it is only legitimate to perform the rotating-wave approximation in the interaction with the mechanical bath after having introduced polariton eigenmodes \cite{lemonde2013nonlinear}.

\section{The linear regime}\label{Sec: linear}

We first consider $\hat{H}$ in the absence of non-linear terms ($g_{1,2}=0$), when the system is described by non-interacting polaritons, i.e., normal modes which are mixtures of photon and phonon modes. The polaritons are characterized by their own frequencies, damping rates, and temperatures, which are obtained below and allow us to compute the cavity DOS in the linear regime. 

These results will also be useful in Sec.~\ref{Sec: nonlinear}, as the polariton basis is more convenient than the photon-phonon basis for handling the nonlinearity. Although we will only include non-linear effects in a chain with $N=2$ unit cells (like in Fig.~\ref{fig: the system}), we provide in this section the general treatment of arbitrary even $N$, which could be useful for further generalizations.

\subsection{Normal modes and polariton basis}

We focus in the following on $\Delta < 0$, as in the blue-detuned regime ($\Delta >0$) the region of stability is much smaller due to the occurrence of self-induced oscillations \cite{marquardt2006dynamical,ludwig2008optomechanical}. The normal modes are given by $\hat{c}_{\sigma, k}$ polariton operators bringing $\hat{H}_0$ to diagonal form:
\begin{equation}\label{Hamiltonian: linear diag}
\hat{H}_0 = \sum_{\sigma, k} \omega_{\sigma, k} \hat{c}^{\dagger}_{\sigma, k}  \hat{c}_{\sigma, k},
\end{equation}
where $k = 2j\pi/N$ (with $j=1,2,\ldots , N$). All wavevectors are defined ${\rm mod}~2\pi$, and $\sigma = +~(-)$ indicates the upper (lower) polariton branch with frequency:
\begin{equation}\label{polariton frequencies}
\omega_{\pm, k} =  \sqrt{\frac{\Delta^2 + \omega^2_m}{2} \pm \frac12 \sqrt{(\Delta^2 - \omega^2_m)^2 + 16 \omega_m |\Delta|  G^2_k }}.
\end{equation}
$G_k$ is the $k$-dependent many-photon coupling:
\begin{equation}\label{G_k}
G_k = \sqrt{ G^2_1+G^2_2+2G_1 G_2 \cos k },
\end{equation}
which can be alternatively defined from $G_1+G_2 e^{ik} = G_k e^{i \phi_k}$. The phase $\phi_k$ is irrelevant for the polariton frequencies, but enters the linear transformation leading to Eq.~(\ref{Hamiltonian: linear diag}):
\begin{equation}\label{linear_transf}
( \hat{b}_n~\hat{d}_n )^{T} =  \sum_{k} \frac{e^{ikn}}{\sqrt{N}} ~V(k) (\hat{c}_{-, k}~\hat{c}_{+, k}~\hat{c}^{\dagger}_{-, -k}~\hat{c}^{\dagger}_{+, -k} )^{T},
\end{equation}
where $T$ indicates the transpose. The matrix elements of $V(k)$ are explicitly given as follows:
\begin{align}
&V_{1,2\mp 1}(k)= \frac{e^{i\phi _{k}/2}  }{2} f_\pm \left(\omega_m/\omega _{-,k}\right)
\cos\theta_k,  \label{Vmatrix1} \\
&V_{1,3\mp 1}(k)=\frac{e^{i\phi _{k}/2}  }{2} f_\pm \left(\omega_m/\omega _{+,k}\right)
\sin\theta_k, \label{Vmatrix2} \\
&V_{2,2\mp 1}(k)=- \frac{e^{-i\phi _{k}/2}  }{2} f_\pm \left(|\Delta|/\omega _{-,k}\right)
\sin\theta_k, \label{Vmatrix3} \\
&V_{2,3\mp 1}(k)=\frac{e^{-i\phi _{k}/2}  }{2} f_\pm \left(|\Delta|/\omega _{+,k}\right)
\cos\theta_k, \label{Vmatrix4} 
\end{align}
where we defined $f_{\pm}(x)= \sqrt{x} \pm \sqrt{1/x}$. The mixing angle $\theta_k$ satisfies: 
\begin{equation} \label{theta_k}
\tan 2\theta_k = \frac{4 G_k \sqrt{|\Delta| \omega_m} }{\Delta^2 - \omega^2_m},~~~\theta_k \in [0, \pi/2].
\end{equation}

If we specialize these expressions to $N=2$, the wavevector can only assume the values $k=\pi, 2\pi$. Therefore, $k$ and $-k$ [appearing on the right hand side of Eq.~(\ref{linear_transf})] are always equivalent in our case. For $N=2$, the many-photon couplings are simply given by $G_{2\pi} = G_+$ and $G_{\pi} = G_-$, where
\begin{equation}
G_{\pm} =G_1 \pm G_2 .
\end{equation} 
Without loss of generality we will assume $G_+,G_- \geq 0$ (i.e., $G_1 \geq |G_2| \geq 0$). If we require $\omega_{\pm, k} >0$, we obtain from Eq.~(\ref{polariton frequencies}) an approximate condition for the onset of the optomechanical instability  \cite{dorsel1983optical,lemonde2015real,xu2015quantum}:
\begin{equation} \label{instability}
G_{\pm} <  \frac12 \sqrt{|\Delta|\omega_m} \equiv G_{\rm cri}.
\end{equation}
Here, the critical value $ G_{\rm cri}$ neglects the effect of relatively small dampings $\kappa,\gamma$. Our treatment of non-linear interactions will be only valid sufficiently far from the instability.

\subsection{Polariton damping rates and polariton temperatures}

The effective polariton damping rates and temperatures are derived by substituting Eq.~\eqref{linear_transf} into Eq.~\eqref{H_diss}. We can apply the rotating-wave approximation to the interaction of the system with the mechanical baths, giving: 
\begin{align}\label{dissipation_c_mech}
i \sqrt{\frac{\gamma}{2 \pi \rho_m}} 
\sum_{k,n,j} \frac{e^{ikn}}{\sqrt{N}} 
\Big[(V_{11}(k)+V_{13}(k))  \hat{p}^{\dagger}_{n,j}  \hat{c}_{-,k} \nonumber \\
 + (V_{12}(k)+V_{14}(k))  \hat{p}^{\dagger}_{n,j}  \hat{c}_{+,k} - {\rm H.c.} \Big],
\end{align}
where $n=1,2, \ldots, N$. The matrix elements of $V(k)$ can be found in Eqs.~(\ref{Vmatrix1}--\ref{Vmatrix4}) and satisfy $[V_{ij}(-k)]^*=V_{ij}(k)$. Similarly, the interaction with the optical baths is:
\begin{align}\label{dissipation_c_opt}
i \sqrt{\frac{\kappa}{2 \pi \rho_c}} 
&\sum_{k,n,j} \frac{e^{ikn}}{\sqrt{N}}  
\Big[  \left( V_{21}(k) \hat{o}^{\dagger}_{n,j} - V_{23}(k) \hat{o}_{n,j} \right) \hat{c}_{-,k}\nonumber \\ 
&+ \left( V_{22}(k) \hat{o}^{\dagger}_{n,j} - V_{24}(k) \hat{o}_{n,j} \right) \hat{c}_{+,k}  - {\rm H.c.}\Big].
\end{align}
Here we should not discard the counter-rotating terms since the bath frequencies in Eq.~(\ref{H_diss}) satisfy $\Delta_{n,j} > - \omega_L$, which allows for ``quantum heating'' effects \cite{lemonde2015real}.

\begin{figure}
\centering
\includegraphics[width=\columnwidth]{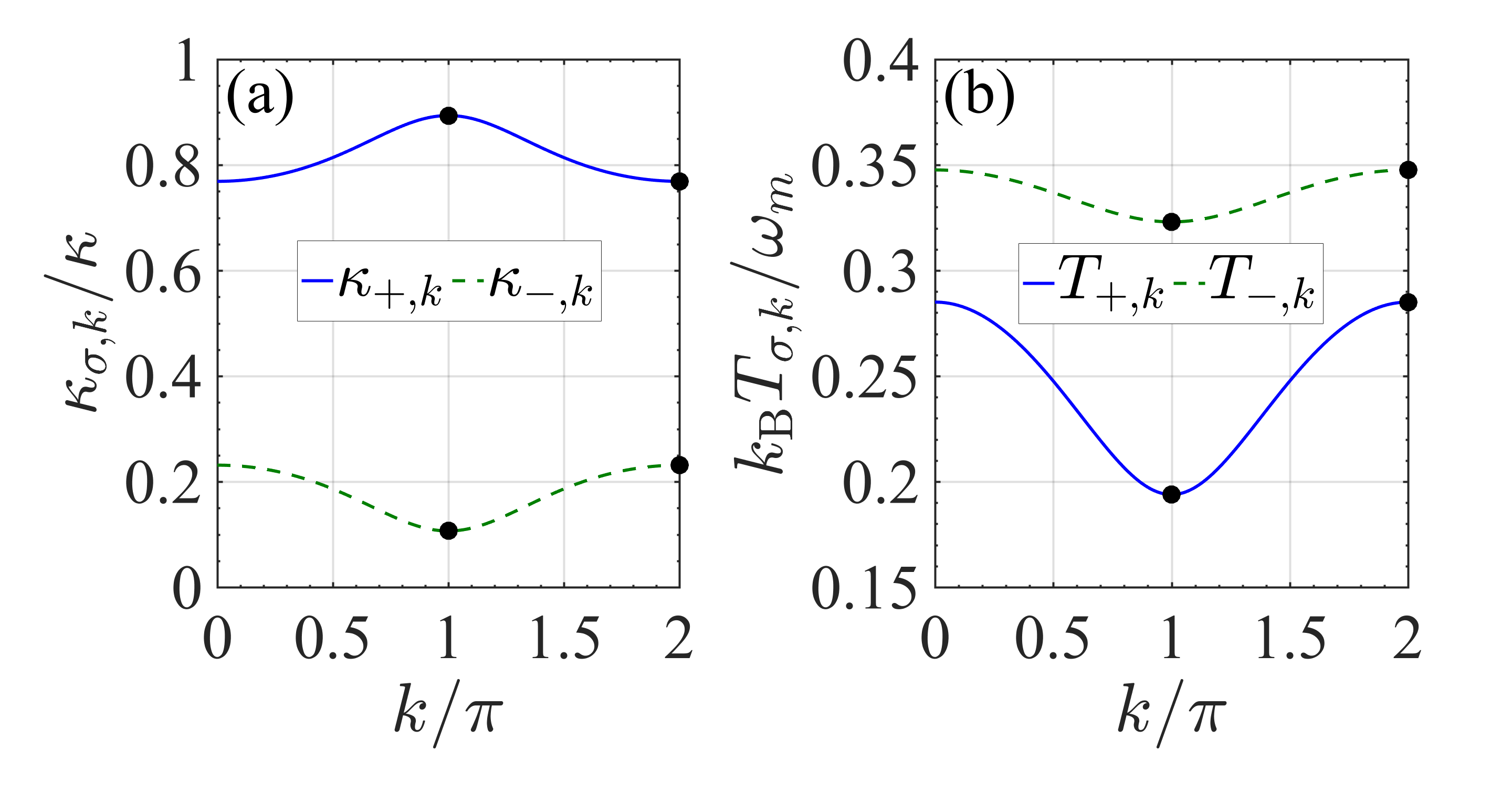}
\caption{(Color online) Polariton damping rates (a) and polariton temperatures (b) as functions of  wave vector $k$. The upper ($+$) and lower ($-$) branches are respectively shown as solid and dashed curves. For $N=2$, only the $\pi$ and $2\pi$ modes (black points) play a role. We used Eqs.~(\ref{dampings}) and (\ref{occupations}) with the parameters $\Delta=-1.5\omega_m$, $G_{-}=0.2\omega_m$, $G_{+}=0.4\omega_m$, $T=0$, and $\gamma/\kappa = 10^{-4}$.}
\label{fig: damping temperature}
\end{figure}

Equations~(\ref{dissipation_c_mech}) and (\ref{dissipation_c_opt}) show that each of the polaritons, which is in general a linear combination of all mechanical and optical modes, interacts with $2N$ reservoirs. Treating all the polariton baths as independent, which is justified if in the assumption of weak dissipation $\kappa, \gamma \ll \omega_{\sigma,k}, \vert \omega_{+,k} - \omega_{-,k} \vert$,  we can perform a standard derivation of the Heisenberg-Langevin equations for $\hat{c}_{\sigma, k}$:
\begin{equation} \label{eq: A_k}
\dot{\hat{c}}_{\sigma, k}(t)  = -i \omega_{\sigma,k} \hat{c}_{\sigma, k}(t)  -\frac{\kappa_{\sigma,k}}{2}  \hat{c}_{\sigma, k}(t) - \sqrt{\kappa_{\sigma,k}}  \hat{\xi}_{\sigma, k}(t), 
\end{equation}
where the polariton damping rates are given by: 
\begin{align}\label{dampings}
\kappa_{+,k} = \kappa \left[ \vert V_{22}(k) \vert ^2 - \vert V_{24}(k) \vert ^2 \right] + \gamma  \vert V_{12}(k)+V_{14}(k) \vert ^2, \nonumber\\
\kappa_{-,k} = \kappa \left[ \vert V_{21}(k) \vert ^2 - \vert V_{23}(k) \vert ^2 \right] + \gamma  \vert V_{11}(k)+V_{13}(k) \vert ^2. 
\end{align}
The noise operators $\hat{\xi}_{\sigma, k}(t)$ satisfy:
\begin{equation}
\langle \hat{\xi}_{\sigma, k}^{\dagger}(t_1) \hat{\xi}_{\sigma, k}(t_2) \rangle = n_{\sigma, k} ~ \delta(t_1 - t_2),
\end{equation}
with the following occupation numbers of the polariton modes:
\begin{align}\label{occupations}
n_{+, k} = \frac{\kappa \vert V_{24}(k) \vert^2 + \gamma \vert V_{12}(k) + V_{14}(k) \vert^2 n(\omega_{+, k})}{\kappa_{+, k}} , \nonumber \\
n_{-, k} = \frac{\kappa \vert V_{23}(k) \vert^2 + \gamma \vert V_{11}(k) + V_{13}(k) \vert^2 n(\omega_{-, k}) }{\kappa_{-, k}} .
\end{align}
In the equations above, which generalized the two-mode case \cite{lemonde2013nonlinear} to our periodic chain, $n(\omega_{\sigma, k})=[e^{\omega_{\sigma, k}/(k_B T)}-1 ]^{-1}$ is the Bose-Einstein distribution evaluated at polariton frequency $\omega_{\sigma, k}$ and environmental temperature $T$. As expected, the stationary state is out of equilibrium: interpreting the polariton occupations in terms of effective temperatures $T_{\sigma, k}$ (through the relation $n_{\sigma, k} =[e^{\omega_{\sigma, k}/(k_B T_{\sigma, k})}-1 ]^{-1}$), it is easily seen that the four polariton temperatures are in general all different, and do not coincide with the environmental temperature $T$. An example of the momentum dependence of the damping rates and effective temperatures is given in Fig.~\ref{fig: damping temperature}.

Using Eqs.~(\ref{Vmatrix1}-\ref{Vmatrix4}), we find a more explicit expression for the damping rates:
\begin{align} \label{kappa2}
&\kappa_{+,k} = \kappa \cos^2 \theta_{k} + \gamma  \frac{\omega_m}{\omega_{+,k}} \sin^2 \theta_{k},\nonumber \\
&\kappa_{-,k} = \kappa \sin^2 \theta_{k} + \gamma  \frac{\omega_m}{\omega_{-,k}} \cos^2 \theta_{k},
\end{align}
which makes clear how the mixing angle $\theta_k$ plays an important role in controlling the decay of the polariton modes. In particular, for $\Delta=-\omega_m$ one has $\theta_k = \pi/4$, implying strongly mixed polariton modes and $\kappa_{+, k}\simeq \kappa_{-,k} \simeq \kappa/2$ (assuming, as is usually the case, that the contribution from $\gamma \ll \kappa$ is negligible). On the other hand, for $\vert \Delta \vert$ away from $\omega_m$, the $+~(-)$ polaritons have a photonic (phononic) character. We shall be especially interested  in the limit $\vert \Delta \vert \gg \omega_m$; in this regime, $\theta_k \ll 1 $ and the phononic mode $\hat{c}_{-,k}$ is characterized by $\kappa_{-,k} \ll \kappa$. 

For the polariton occupations, the expressions analogous to Eq.~(\ref{kappa2}) can also  be easily obtained:
\begin{align} \label{occuptions2}
&n_{+,k} = \frac{\left(|\Delta|-\omega_{+,k}\right)^2}{4|\Delta|\omega_{+,k}}\left( 1 +\frac{\gamma}{\kappa}  \frac{\omega_m}{\omega_{+,k}} \tan^2 \theta_{k}\right)^{-1},\nonumber \\
&n_{-,k} = \frac{\left(|\Delta|-\omega_{-,k}\right)^2}{4|\Delta|\omega_{-,k}}\left( 1 +\frac{\gamma}{\kappa}  \frac{\omega_m}{\omega_{-,k}} \cot^2 \theta_{k}\right)^{-1},
\end{align}
where for simplicity we assumed $T=0$.

\subsection{Cavity DOS}

In the linear regime, physical observables can be directly obtained from the polariton representation. In the following we will pay special attention to the cavity DOS, which is obtained as follows:
\begin{equation}\label{formula: DOS}
\rho^{(0)}_d [\omega] = -\frac{1}{\pi} {\rm Im} [\mathcal{G}_{0}^R (\hat{d}_n, \hat{d}^{\dagger}_n; \omega) ],
\end{equation}
where $\mathcal{G}_{0}^R (\hat{A}, \hat{B}; \omega) = -i \int^{\infty}_{-\infty}dt e^{i\omega t} \Theta(t) \langle [\hat{A}(t), \hat{B}(0) ] \rangle$ is the retarded Green's function, with $\Theta(t)$ the Heaviside step function. Note that we use ``$0$" to distinguish the linear case from the nonlinear one, thus the time evolution and expectation value refer here to the linear Hamiltonian ${\hat{H}_{\rm 0} + \hat{H}_{\rm diss}}$.

Since all the cavities of the ring are equivalent, $\rho^{(0)}_d [\omega]$ is independent of $n$. This is easily seen by computing $\mathcal{G}_0^R (\hat{d}_n, \hat{d}^{\dagger}_n; \omega)$ in terms of polariton modes. The polariton Green's functions are diagonal and have no anomalous components, i.e., $\mathcal{G}_0^R (\hat{c}_{\sigma,k}, \hat{c}_{\sigma',k'}; \omega) = 0$ and $\mathcal{G}_0^R (\hat{c}_{\sigma,k}, \hat{c}^\dag_{\sigma',k'}; \omega)=\delta_{\sigma\sigma'}\delta_{kk'}\mathcal{G}_0^R(\sigma,k;\omega)$ where:
\begin{equation}\label{GF_polaritons}
\mathcal{G}_0^R (\sigma,k; \omega ) = \left(\omega - \omega_{\sigma, k} + i \kappa_{\sigma, k} /2\right)^{-1}.
\end{equation}
The linear transformation Eq.~(\ref{linear_transf}) gives:
\begin{align}\label{GR_cavity}
&\mathcal{G}_0^R (d_n, d^{\dagger}_n; \omega) = \frac{1}{N}  \sum_{k} \left[|V_{21}(k)|^2 \mathcal{G}_0^R (-,k; \omega ) \right. \nonumber \\
& +|V_{22}(k)|^2 \mathcal{G}_0^R (+,k; \omega ) +|V_{23}(k)|^2 (\mathcal{G}_0^R (-,k; -\omega ))^* \nonumber \\
& \left. +|V_{24}(k)|^2 (\mathcal{G}_0^R (+,k; -\omega ))^*\right],
\end{align}
which allows us to compute the cavity DOS of the optomechanical ring in the linear regime. We will focus on the case $N=2$ and consider the much larger features at $\omega>0$, consisting in four peaks at the normal mode frequencies. The strength of these peaks is given by $|V_{2j}(k)|^2 $ ($j=1,2$ and $k=\pi,2\pi$) and the width is the corresponding polariton damping rate.

\section{Treatment of nonlinear interactions}\label{Sec: nonlinear}

To take into account the non-linear interactions, it is more convenient to express Eq.~(\ref{Hamiltonian: nonlinear}) in terms of polaritons:
\begin{align}\label{nonlinear_int_c}
\hat{H}_{\rm nl} = & \sum_{s} 
\left(\tilde{g}_s \hat{c}^{\dagger}_{\sigma, k}  \hat{c}_{\mu, q} \hat{c}_{\nu, k-q}+
\tilde{g}'_s \hat{c}^{\dagger}_{\sigma, k}  \hat{c}^{\dagger}_{\mu, q}\hat{c}^{\dagger}_{\nu, -k-q}  +{\rm H.c.}\right) \nonumber \\
& +\sum_{\sigma, k} A_{\sigma,k} \left(\hat{c}^{\dagger}_{\sigma, k}+\hat{c}_{\sigma, k} \right),
\end{align}
where $s \equiv(\sigma,\mu,\nu;k,q)$ is a collective label for the scattering process. In our specific model  momentum is conserved mod $2\pi$ but, besides this restriction, all possible scattering terms are allowed in Eq.~(\ref{nonlinear_int_c}). The coefficients $\tilde{g}_s,\tilde{g}'_s$ are proportional to $g_{1,2}$ and can be evaluated by using Eq.~(\ref{linear_transf}), although the explicit expressions are rather cumbersome (see Appendix~\ref{appendix1}). The linear terms in the second line arise from normal ordering, with $A_{\sigma,k}$ also proportional to $g_{1,2}$.

Although there are many nonlinear interaction terms in Eq.~(\ref{nonlinear_int_c}), only few of them play a significant role.  In fact, due to the smallness of the nonlinear couplings $g_{1,2}$, it is allowed to neglect in Eq.~(\ref{nonlinear_int_c}) all but the resonant processes. This leaves us with:
\begin{equation}\label{nonlinear_int_resonant}
\hat{H}_{\rm nl} \simeq  {\sum_{s}}^\prime
\left(\tilde{g}_s c^{\dagger}_{\sigma, k}  \hat{c}_{\mu, q} \hat{c}_{\nu, k-q}
  +{\rm H.c.}\right),
\end{equation}
where the prime indicates that we only include the terms which can satisfy:
\begin{equation}\label{resonance}
\omega_{\sigma, k} =  \omega_{\mu, q} + \omega_{\nu, k-q},
\end{equation}
for a suitable choice of parameters. The conditions to realize such resonant scattering processes are discussed in detail below. Based on the simplified non-linear interaction Eq.~(\ref{nonlinear_int_resonant}), it is possible to compute the full retarded Green's functions $\mathcal{G}_R$ and identify the regimes where effects of the non-linear interaction are strongest.

\subsection{Resonant conditions}\label{sec_resonances}

\begin{figure}
\centering
\includegraphics[width=\columnwidth]{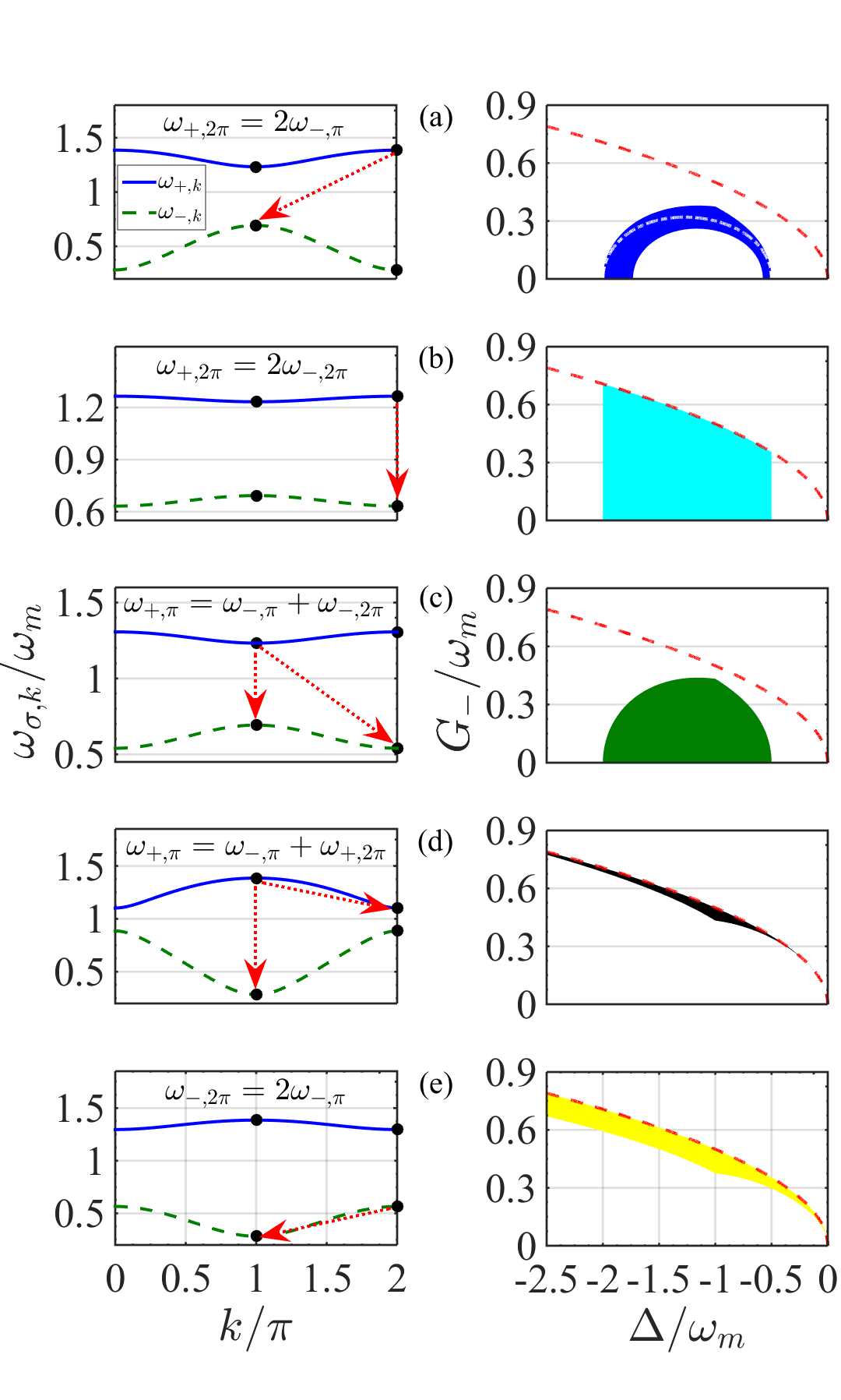}
\caption{(Color online) The five allowed resonant scattering processes. For each case, the left panel illustrates the resonant scattering between polaritons. The corresponding right panel gives the allowed region (in shaded color). In the left panels, the two polariton branches are given by Eq.~(\ref{polariton frequencies}). We used $\Delta = -\omega_m$, $G_- = 0.26 \omega_m$ (a,b,c), $G_- = 0.46 \omega_m$ (d,e), and the value of $G_+$ enforcing resonance. In the right panels we have indicated the instability line $G_-= \sqrt{|\Delta| \omega_m}/2 $ (red dashed). The top right panel also shows the resonant line of the two-mode system (white dot-dashed).}
\label{fig: regime_resonant}
\end{figure}

Equation~(\ref{resonance}) can be interpreted in terms of a decay process of polariton $(\sigma, k)$ into two lower-energy polaritons. Five of these processes are allowed, and we have illustrated them in Fig.~\ref{fig: regime_resonant}. 

For a generic optomechanical chain, Eq.~(\ref{resonance}) is not satisfied for any of the five processes, but we can try to enforce one of the resonant conditions by tuning the controllable parameters $G_\pm$ and $\Delta$. In the following discussion we choose to adjust $G_+$, which allows us an easier comparison to the two-mode system. Then, at given values of $\Delta$ and $G_-$, Eq.~(\ref{resonance}) determines the resonant value of $G_+$. The only difficulty is that sometimes such solution does not exist. One should still pay attention to the stability condition Eq.~(\ref{instability}), and the specific form of the dispersion relation Eq.~(\ref{polariton frequencies}) gives further restrictions on the parameter range where resonances can occur. Therefore, in each of the five cases we have indicated in Fig.~\ref{fig: regime_resonant} the colored region where a physical solution for $G_+$ can actually be found (right panels). 

To give an explicit example, we specialize on process (e), which will play a special role in the following. The (yellow) allowed region of  Fig.~\ref{fig: regime_resonant}(e) is defined by:
\begin{equation}
G_{-, {\rm min}}^{(e)} \leq G_- \leq G_{\rm cri},
\end{equation}
where
\begin{equation} \label{Glower_e}
G_{-,{\rm min}}^{(e)}  =
 \left \{
\begin{array} {ll} 
\sqrt{\frac{3\omega_m}{16\vert\Delta\vert }\left(\Delta^2-\frac{\omega_m^2}{4}\right)}
,  ~& \Delta < -\omega_m,  \\
\sqrt{\frac{3\vert\Delta\vert }{16\omega_m}\left(\omega_m^2-\frac{\Delta^2}{4}\right)}
 , ~&  -\omega_m \leqslant \Delta \leqslant 0. 
\end{array}
\right.
\end{equation}
Within this region, the resonance is enforced by the following choice of $G_+$:
\begin{equation}\label{Gplus_e}
G_+^{(e)}=2\sqrt{\frac{(\omega^2_{-,\pi}-\omega^2_m/4)(\omega^2_{-,\pi}-\Delta^2/4)}{|\Delta|\omega_m}}.
\end{equation}
The remaining four cases are given in Appendix~\ref{appendix2}. We also note that, in fact, there are two more processes allowed by energy and momentum conservation:
\begin{equation}
\omega_{+, 2\pi} = 2 \omega_{+, \pi},~~~{\rm and}~~~\omega_{+, 2\pi} =\omega_{+, \pi}+ \omega_{-, \pi}.
\end{equation} 
However, a detailed analysis shows that these processes have vanishing allowed regions, thus are omitted in Fig.~\ref{fig: regime_resonant}. 

In general, even if the allowed regions intersect, different resonances cannot be realized simultaneously. For example, the (cyan) region of resonance (b) contains the (green) region of resonance (c) but in general $G_+^{(b)} \neq G_+^{(c)}$ (see Appendix~\ref{appendix2}). The only exception is at $G_+=G_-$, when processes (a), (b), and (c) are all simultaneously realized. This occurs because (a), (b), and (c) are all inter-branch resonances, where the $+$ polariton exclusively decays into the $-$ branch, and at this special point the chain breaks down into two disjoint two-mode systems.  

On the other hand, processes (d) and (e) have an intra-branch character, since the initial excitation decays into respectively one or two polaritons belonging to the initial branch. An important feature of these intra-branch resonances is that they can be realized in an unbounded region, i.e., the value of $|\Delta|$ can become arbitrary large at the expense of increasing the dressed coupling $G_{-}$. This feature is new with respect to the two-mode case \cite{lemonde2013nonlinear,borkje2013signatures}, and is actually very useful to achieve stronger nonlinear effects. 

Another interesting property of these intra-branch processes, shared with (b), is that they allow to explore the resonance condition in the proximity of the instability regime, where non-linear effects are expected to be naturally enhanced. This is impossible in the two-mode case, where instability boundary and resonant condition are always well distinct (see the dashed and dot-dashed lines in the top right panel of Fig.~\ref{fig: regime_resonant}). 

\subsection{Retarded self-energy}

To characterize the leading corrections to the non-interacting retarded Green's function we make use of the Keldysh diagrammatic technique, which is appropriate for the driven-dissipative system at hand. 

As we restrict ourselves to resonant processes, we compute $\mathcal{G}^R(\sigma,k;\omega) \equiv \mathcal{G}^R (c_{\sigma,k}, c^\dag_{\sigma,k}; \omega)$ by considering the approximate nonlinear interaction Eq.~(\ref{nonlinear_int_resonant}). It is well known that the Dyson equation for the retarded Green's function is diagonal in the $R$ index, and Eq.~(\ref{GF_polaritons}) becomes:
\begin{align}\label{GF_general}
\mathcal{G}^R (\sigma, \kappa; \omega ) = \left(\omega - \omega_{\sigma, k} + i \kappa_{\sigma, k} /2 - \Sigma^R_{\sigma, k}(\omega) \right)^{-1}.
\end{align}
However, the retarded self-energy $\Sigma^R_{\sigma, k}(\omega)$ involves the advanced ($A$) and Keldysh ($K$) Green's functions. In the noninteracting case, they are obtained from Eqs.~(\ref{GF_polaritons}) and (\ref{occupations}) as follows:
\begin{align}
\mathcal{G}_0^A(\sigma, \kappa; \omega ) &=\left[ \mathcal{G}_0^R(\sigma, \kappa; \omega ) \right]^*, \\
\mathcal{G}_0^K(\sigma, \kappa; \omega ) & =2i (2n_{\sigma,\kappa}+1){\rm Im} \left[ \mathcal{G}_0^R(\sigma, \kappa; \omega ) \right].
\end{align}

\begin{figure}
\centering
\includegraphics[width=\columnwidth]{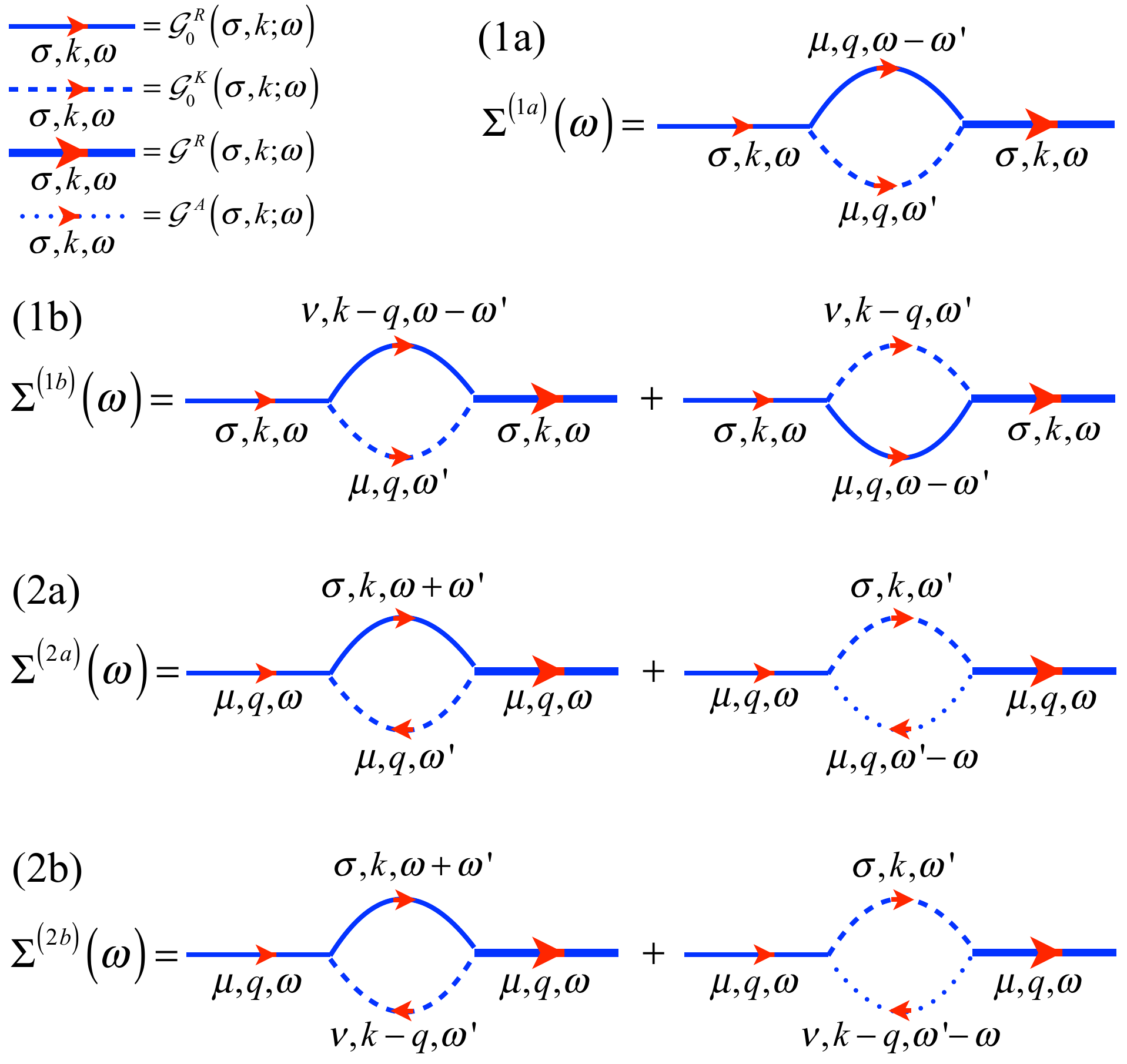}
\caption{(Color online) The four types of second-order self-energy, respectively given by Eqs.~\eqref{sigma:1a}, \eqref{sigma:1b}, \eqref{sigma:2a}, and \eqref{sigma:2b} of the main text (from top to bottom). The labeling of the Green's functions with $(\sigma,k)$, $(\mu,q)$, and $(\nu,k-q)$  follows the notation of Eq.~(\ref{nonlinear_int_resonant}), taking into account that (1a) and (2a) involve identical polaritons, thus $(\nu,k-q) = (\mu,q)$.} 
\label{fig: self-energy diagram}
\end{figure}

To second-order, the result of the Keldysh calculation is illustrated by the self-energy diagrams shown in Fig.~\ref{fig: self-energy diagram}. Below we will give the explicit expression of each diagram, see Eqs.~(\ref{sigma:1a}-\ref{sigma:2b}). As a preliminary discussion, we first describe the structure of the different contributions originating from a given scattering term, of the form $\tilde{g}_s \hat{c}^{\dagger}_{\sigma,k} \hat{c}_{\mu,q} \hat{c}_{\nu,k-q}$. In fact, such a nonlinear interaction gives rise to two types of self-energies. 

In the first type [diagrams (1a) and (1b) of Fig.~\ref{fig: self-energy diagram}], the incoming polariton $(\sigma,k)$ decays to $(\mu, q) $ and $(\nu, k-q )$. These polaritons then recombine into the outgoing $(\sigma,k)$ Green's function. The second type of self-energy occurs when, say, the incoming polariton is $(\mu, q) $ [diagrams (2a) and (2b) of Fig.~\ref{fig: self-energy diagram}]. The simultaneous destruction of $(\mu, q) $ and $(\nu, k-q )$ generates a $(\sigma,k)$ polariton in the internal bubble.

While the above considerations are the same as the two-mode system \cite{lemonde2015real,lemonde2013nonlinear}, a difference arises here due to the presence of four modes instead of two. In the two-mode system, the two low-energy polaritons are necessarily identical but here they can belong to two different modes, as shown in Fig.~\ref{fig: regime_resonant}(c) and (d). Correspondingly, we distinguish in Fig.~\ref{fig: self-energy diagram} between self-energies of type (1a) and (2a), which involve identical polaritons, and type (1b) and (2b), where the low-energy polaritons are different. As we discuss below, the difference between having identical/unequal polaritons is just a symmetry factor.

Having clarified the general structure of the different contributions to the self-energy, we write it as follows:
\begin{equation}\label{sigmaR_sum}
\Sigma^R_{\sigma,k}(\omega) = \sum_{i } ~\Sigma^{(i)}_{\sigma,k}(\omega),
\end{equation}
where $i$ runs over the five scattering processes of Fig.~\ref{fig: regime_resonant}. In general, some of the $\Sigma^{(i)}_{\sigma,k}(\omega)$ will be absent from the sum since $(\sigma,k)$ is not involved in all the processes. For example, polariton $(+,2\pi)$ only appears in (a), (b), and (d). It is also clear that $(i)$ and $(\sigma,k)$ determine unambiguously if the diagram is of type (1a), (1b), (2a), or (2b). The complete classification is given in Table~\ref{table: self-energy}.

Finally, we give the explicit expression of $\Sigma^{(i)}_{\sigma,k}(\omega)$. If it belongs to type (1a), it is given by:
\begin{align}\label{sigma:1a}
\Sigma^{(i)}_{\sigma, k}\left( \omega \right) &= 2i \tilde{g}_s^2 \int\limits^{\infty}_{-\infty} \frac{d \omega'}{2\pi} \mathcal{G}^K_0 (\mu, q; \omega' ) \mathcal{G}^R_0 (\mu, q; \omega -\omega' ) \nonumber\\
&= \frac{2 \tilde{g}_s^2 \left(1+2 n_{\mu,q}\right) }{\omega - 2\omega_{\mu,q} + i \kappa_{\mu,q} } , \qquad \mathrm{[type~(1a)]}
\end{align}
where $s$ is directly related to $(i)$. Instead, if $\Sigma^{(i)}_{\sigma,k}(\omega)$ is of type (1b) it is given by:
\begin{align}\label{sigma:1b}
 \Sigma^{(i)}_{\sigma,k}(\omega) =&  \frac{i \tilde{g}_s^2}{2}  \int\limits^{\infty}_{-\infty} \frac{d \omega'}{2\pi} \left[ \mathcal{G}^K_0 (\mu,q; \omega' ) \mathcal{G}^R_0 (\nu,k-q; \omega-\omega' )\right.
\nonumber\\
  & \qquad \quad  +\left. \mathcal{G}^K_0 (\nu,k-q; \omega' ) \mathcal{G}^R_0 (\mu,q; \omega-\omega' ) \right] \nonumber\\
=&   \frac{ \tilde{g}_s^2 (1 + n_{\mu,q}+ n_{\nu,k-q}) }{\omega - \omega_{\mu,q} -\omega_{\nu,k-q} + i (\kappa_{\mu,q}+\kappa_{\nu,k-q})/2 } . \nonumber \\ 
& \hspace{4.1cm} \mathrm{[type~(1b)]}
\end{align}
Type (2a) is given by:
\begin{align}\label{sigma:2a}
\Sigma^{(i)}_{\mu,q}\left( \omega \right) 
=& 2i \tilde{g}_s^2 \int\limits^{\infty}_{-\infty} \frac{d \omega'}{2 \pi} \left[\mathcal{G}^K_0 (\mu,q; \omega' ) \mathcal{G}^R_0 (\sigma,k; \omega'+\omega ) \right. \nonumber\\
&  \qquad \quad ~~~~ \left.+ \mathcal{G}^K_0 (\sigma,k; \omega' ) \mathcal{G}^A_0 (\mu,q; \omega'-\omega)   \right] \nonumber\\
=& \frac{ 4\tilde{g}_s^2 (n_{\mu,q} - n_{\sigma,k}) }{\omega - \omega_{\sigma,k}+\omega_{\mu,q}+ i (\kappa_{\sigma,k}+\kappa_{\mu,q})/2 } ,  \nonumber \\
&\hspace{4.1cm} {\rm [type~(2a)]}
\end{align}
while type (2b) reads:
\begin{align}\label{sigma:2b}
\Sigma^{(i)}_{\mu,q}\left(\omega\right) 
=& \frac{i \tilde{g}_s^2}{2}  \int\limits^{\infty}_{-\infty} \frac{d \omega'}{2 \pi} 
\left[\mathcal{G}^K_0 (\nu,k-q; \omega' ) \mathcal{G}^R_0 (\sigma,k; \omega'+\omega ) \right. \nonumber\\
&  \quad \quad ~~~~  \left.+ \mathcal{G}^K_0 (\sigma,k; \omega' ) \mathcal{G}^A_0 (\nu,k-q; \omega'-\omega)   \right] \nonumber\\
=&\frac{ \tilde{g}_s^2 (n_{\nu,k-q} - n_{\sigma,k}) }{\omega - \omega_{\sigma,k}+\omega_{\nu,k-q} + i (\kappa_{\sigma,k}+\kappa_{\nu,k-q})/2 } . \nonumber \\
& \hspace{4.1cm}  {\rm [type~(2b)]}
\end{align} 

Equations~(\ref{sigma:1a}) and (\ref{sigma:2a}), are in agreement with the ones obtained for two-mode case \cite{lemonde2015real,lemonde2013nonlinear}. The self-energy in Eq.~(\ref{sigma:1b}) is directly related to Eq.~(\ref{sigma:1a}): by setting $(\nu,q-k) = (\mu,q)$ in Eq.~(\ref{sigma:1b}) we recover one half of Eq.~(\ref{sigma:1a}). The additional factor of 2 reflects two equivalent ways to pair the bosonic operators in the self-energy, if the two internal lines belong to identical polaritons. Similarly, setting $(\nu,q-k) = (\mu,q)$ in Eq.~(\ref{sigma:2b}) we recover one quarter of Eq.~(\ref{sigma:2a}). The factor of 4 arises from the two equivalent choices to contract each external bosonic operator with the vertex, which carries two identical polariton operators.

\begin{table}
\centering
\includegraphics[width=0.95\columnwidth]{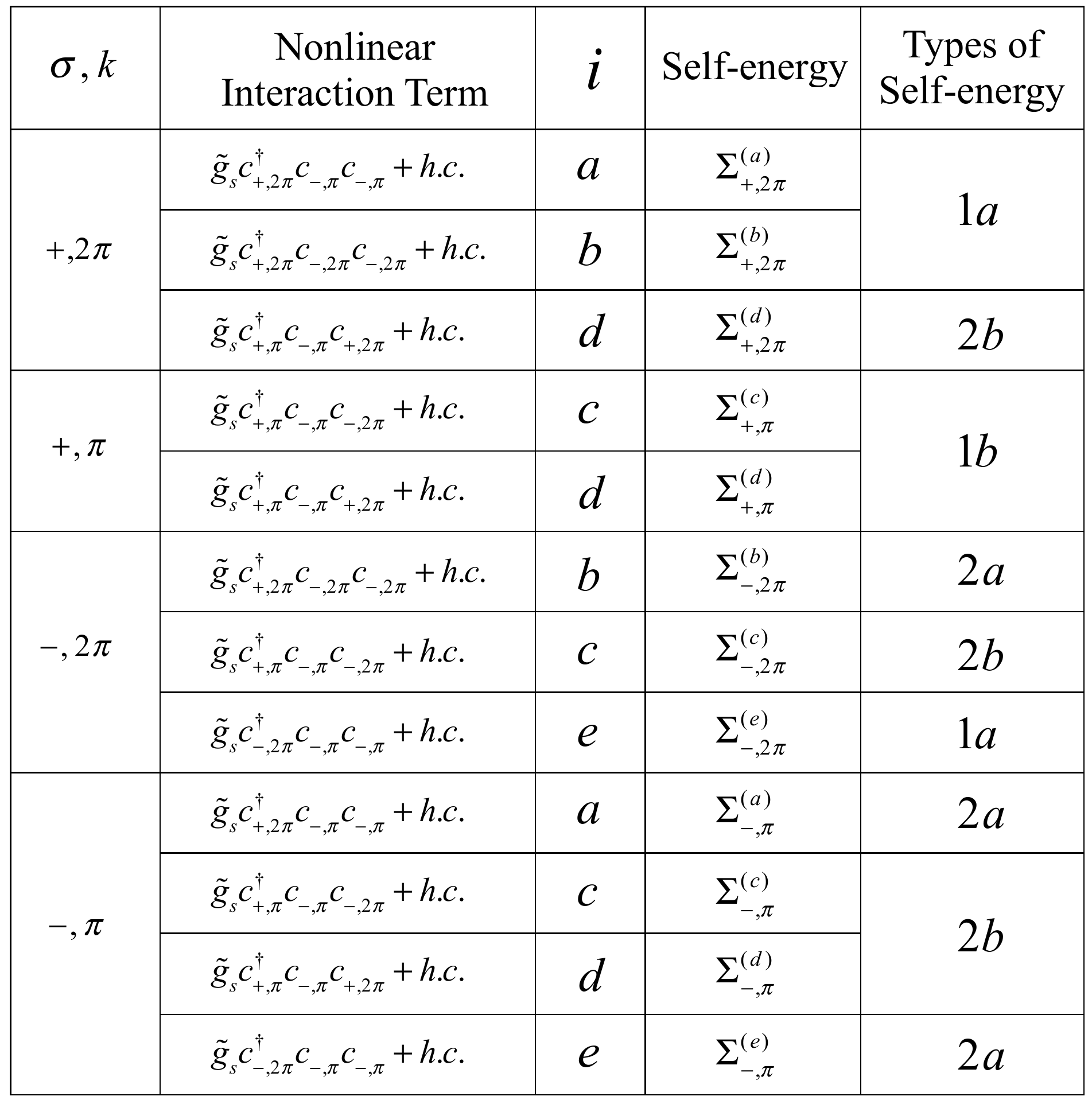}
\caption{Classification of the terms entering Eq.~(\ref{sigmaR_sum}). The last column follows Fig.~\ref{fig: self-energy diagram}.}
\label{table: self-energy}
\end{table}

The above expressions give us the four polariton retarded Green's functions, from which  the cavity Green's function $\mathcal{G}^R(d_n,d^{\dagger}_n;\omega)$ is computed through a transformation identical to Eq.~(\ref{GR_cavity}). Evidently, the cavity DOS: 
\begin{equation}\label{DOS}
\rho_d [\omega] = -\frac{1}{\pi} {\rm Im} [\mathcal{G}^R (\hat{d}_n, \hat{d}^{\dagger}_n; \omega) ],
\end{equation}
is modified from the noninteracting value of Eq.~(\ref{formula: DOS}) by the nonlinear effects. In the next section we will present the physical consequences of the nonlinearity on DOS and OMIT signal, obtained through this formalism. 

\section{Results and discussion} \label{Sec: results and discussions}

\begin{figure}
\centering
\includegraphics[width=\columnwidth]{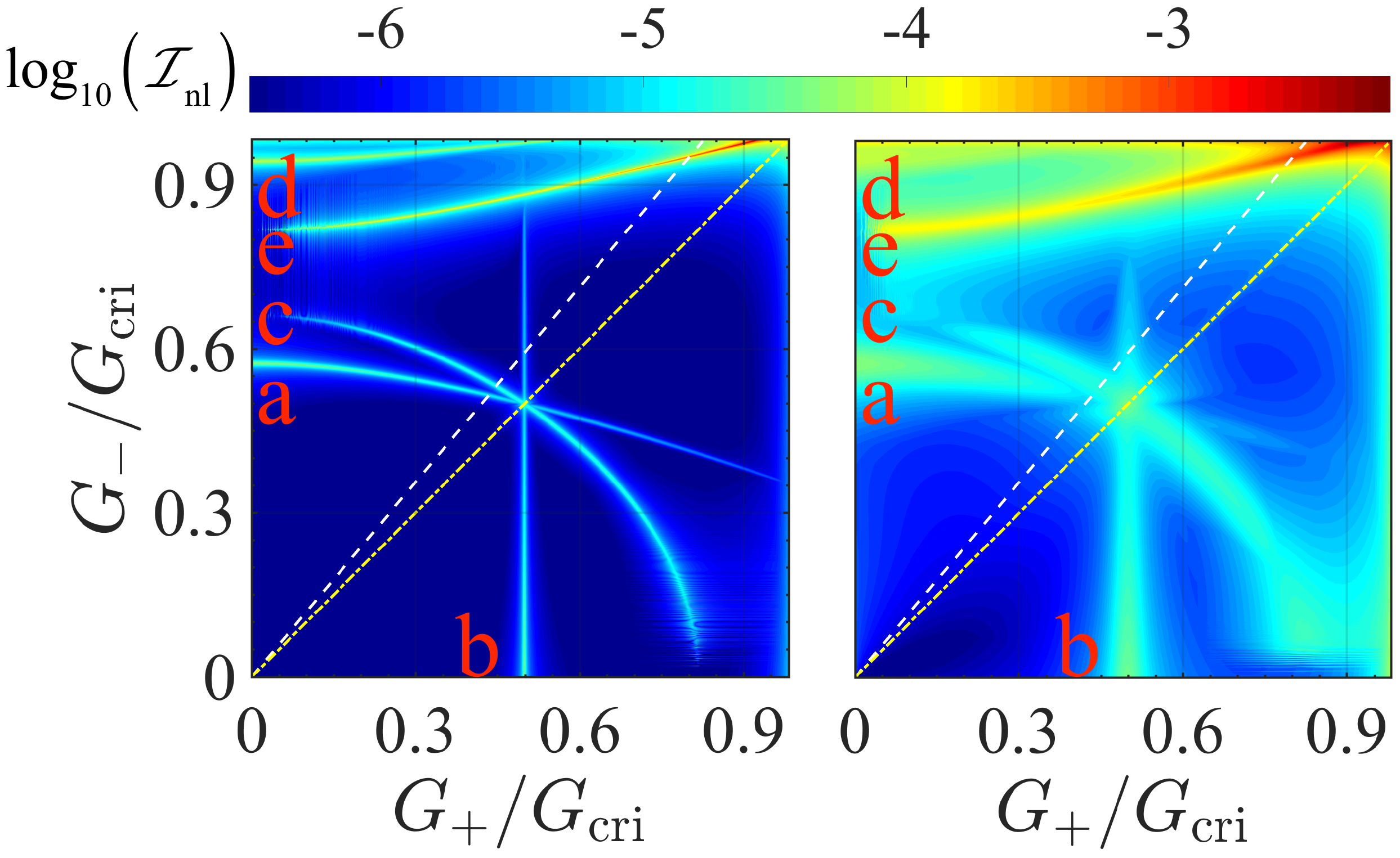}
\caption{(Color online) Plot of $\cal{I}_{\rm nl}$ as a function of $G_{\pm}$ at fixed $\Delta =  -1.5\omega_m$. Nonlinear effects are visible along five curves, corresponding to the resonant conditions of Sec.~\ref{sec_resonances}. Following Fig.~\ref{fig: regime_resonant}, the five cases are labeled as (a,b,c,d,e). In the left panel we assumed a small value $\kappa/\omega_m = 0.01 $, thus the resonances are much sharper. In the right panel we used $\kappa/\omega_m = 0.1$, and the same resonances are still recognizable. In both panels, prominent nonlinear effects are associated with process (e). The (yellow) dot-dashed lines indicate the condition of two decoupled optomechanical cells ($G_+ = G_-$, giving $G_2=0$) while the (white) dashed lines refer to the specific setup discussed in Sec.~\ref{sec: Experimental realization} ($G_-/G_+ =1.187$).  Other parameters are: $\gamma/\kappa=10^{-4}$, $g_1/\kappa=2\times 10^{-3}$, and $T=0$.} 
\label{fig: nonlinearStr_Gp_Gm}
\end{figure}

In this section we would like to identify and discuss the regime where the nonlinearities generate the largest effects, thus we introduce the following quantity: 
\begin{equation} \label{Inl_def}
{\mathcal I}_{\rm nl} = {\rm max}_\omega \left[ \frac{|\rho_{d} (\omega) -  \rho^{(0)}_{d} (\omega)|}{ \rho^{(0)}_d (\omega)} \right],
\end{equation}
which is the largest relative change in the DOS over the whole spectrum. 

A representative plot of ${\mathcal I}_{\rm nl}$ as a function of $G_\pm $ (at fixed detuning $\Delta =- 1.5 \omega_m$) is shown in Fig.~\ref{fig: nonlinearStr_Gp_Gm}, where ${\mathcal I}_{\rm nl}$ was evaluated using the second-order self energy derived with the Keldysh technique. As seen, nonlinear effects are generally negligible except in the regions around the five curves corresponding to the resonant conditions of Fig.~\ref{fig: regime_resonant} (as marked by the labels). Note that ${\mathcal I}_{\rm nl}$ is very small far from the five curves, which provides a justification for neglecting all the nonresonant terms of $\hat{H}_{\rm nl}$.

\begin{figure}
\centering
\includegraphics[width=\columnwidth]{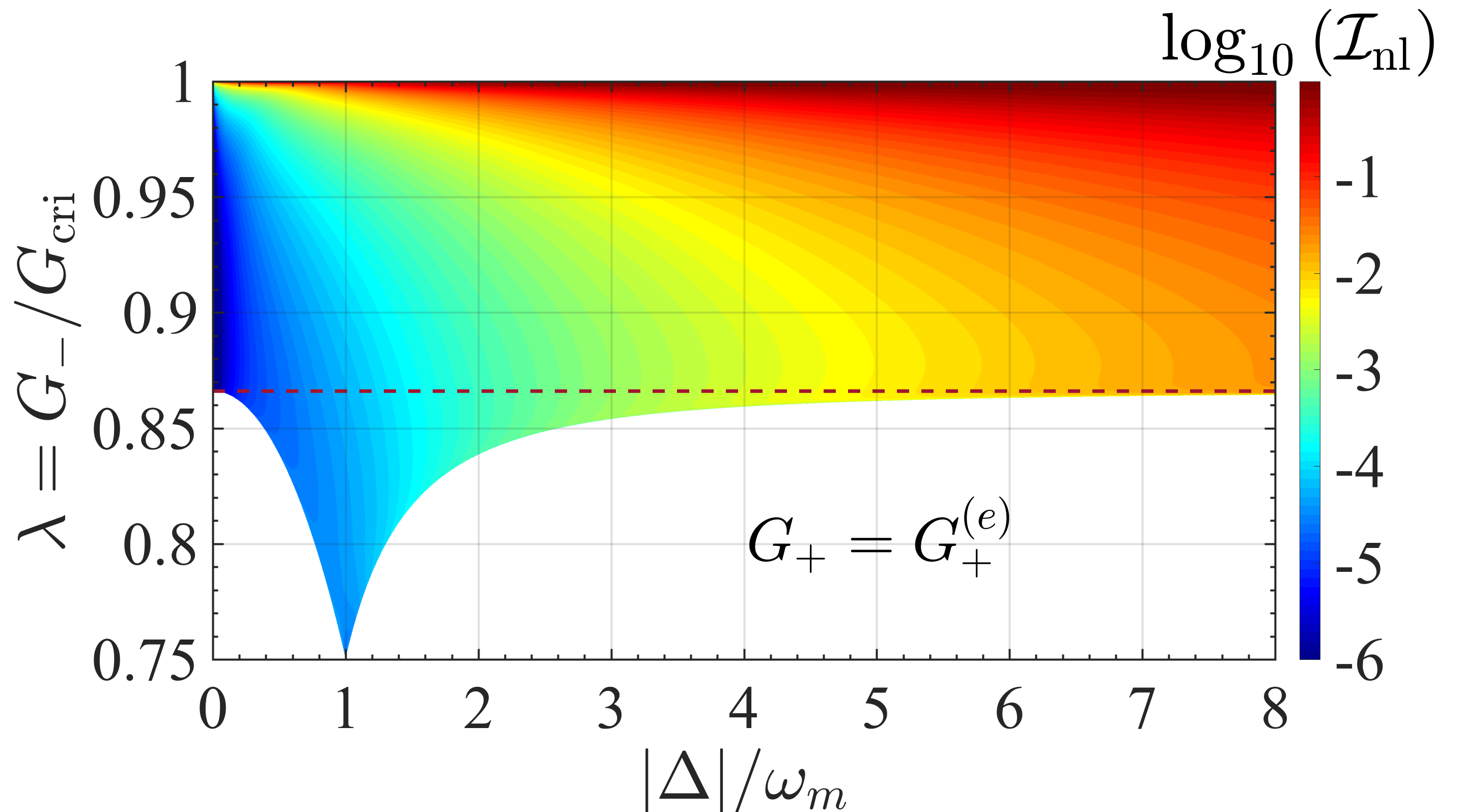}
\caption{(Color online) Plot of $\mathcal{I}_{\rm nl}$ as a function of $|\Delta|/\omega_m$ and $\varg=G_-/G_{\rm cri}$. Here $G_+$ is fixed by resonant condition~(e). The nonlinear signal is only shown in the allowed region of Fig.~\ref{fig: regime_resonant}(e). When $|\Delta|/\omega_m$ is large, the lower edge of the allowed region approaches $\varg=\sqrt{3}/2$ (dashed line).  Other parameters are as in the right panel of Fig.~\ref{fig: nonlinearStr_Gp_Gm}. \color{black}}
\label{fig_I_eresonance}
\end{figure}

At different values of $\Delta$, we obtain features similar to Fig.~\ref{fig: nonlinearStr_Gp_Gm}, with the only notable difference that outside the interval $-2 \omega_m < \Delta<-\omega_m/2$ only the resonant curves (d) and (e) survive, in agreement with the allowed regions shown in Fig.~\ref{fig: regime_resonant}. Accordingly, we suppose that system parameters are tuned to one of the resonances, and focus on process (e). This implies $G_+=G_+^{(e)}$, given in Eq.~(\ref{Gplus_e}). The resulting ${\mathcal I}_{\rm nl}$ becomes a function of the tunable parameters $G_-$ and $\Delta$ and is shown in Fig.~\ref{fig_I_eresonance}. 

From Fig.~\ref{fig_I_eresonance} we see that nonlinear effects can be enhanced by approaching the unstable boundary and working at larger negative detunings. The origin of this behavior will be clarified in the rest of this section, where process~(e) is analyzed in detail.

\subsection{Nonlinear effects on the $(-,2\pi)$ polariton}

The main motivation to choose the purely intra-branch process (e) is that it is the one where the nonlinear effect is strongest, as already seen in Fig.~\ref{fig: nonlinearStr_Gp_Gm}. However, if desired, the following treatment could be adapted to the other four resonances. 

As we will discuss shortly, the peculiar feature of process (e) is that the scattering occurs between very weakly damped `phononic' polaritons. To expose this physics, we derive an analytic expression for ${\mathcal I}_{\rm nl}$ based on:
\begin{equation}\label{R}
C_{\rm eff} \simeq \frac{ i\Sigma^{(e)}_{-, 2\pi}(\omega_{-,2\pi})}{\kappa_{-,2\pi}/2} 
= \frac{4 \tilde{g}_e^2 (1+2n_{-,\pi})}{\kappa_{-,\pi}\kappa_{-,2\pi}},
\end{equation}
which we introduced following the analysis of Ref.~\cite{lemonde2013nonlinear}. The effective coupling is $\tilde{g}_e \equiv \tilde{g}_{(-,-,-;2\pi,\pi)} $ [see the notation introduced after Eq.~(\ref{nonlinear_int_c}) and the formulas of Appendix~\ref{appendix1}] and in the last equality we used Eq.~(\ref{sigma:1a}) for the self-energy. The final form of Eq.~(\ref{R}) is analogous to the two-mode system, since the type of resonant scattering is the same. However, the parameter dependence of the various quantities entering $C_{\rm eff} $ is rather different. We postpone to Sec.~\ref{sec_2mode} an explicit comparison between the two cases.

The connection of $C_{\rm eff} $ and ${\mathcal I}_{\rm nl}$ is easily found by considering the following approximation:
\begin{equation}\label{GR_approx}
\mathcal{G}^R(d_n,d^{\dagger}_n;\omega) \simeq \frac{ |V_{21}(2\pi)|^2/2}{\omega - \omega_{-, 2\pi} + i \kappa_{-, 2\pi} /2 - \Sigma^{(e)}_{-, 2\pi}(\omega) },
\end{equation}
which only includes one contribution to $\mathcal{G}^R(d_n,d^{\dagger}_n;\omega)$ [see Eq.~(\ref{GR_cavity}), where  $\mathcal{G}_0^R(-,2\pi;\omega)$ should be substituted by  $\mathcal{G}^R(-,2\pi;\omega) \equiv \mathcal{G}^R (\hat{c}_{-,2\pi}, \hat{c}^\dag_{-,2\pi}; \omega)$]. This approximation is justified for $\omega \simeq \omega_{-, 2\pi}$, and we have also neglected in Eq~(\ref{GR_approx}) the small contributions to the self-energy from processes (b) and (c). Similar approximations can be applied to the noninteracting Green's function, obtained by setting $\Sigma^{(e)}_{-, 2\pi} \to 0$ in Eq.~(\ref{GR_approx}). Finally, from Eqs.~(\ref{Inl_def}) and~(\ref{GR_approx}), and using the fact that the largest nonlinear effects occur at $\omega = \omega_{-, 2\pi}$, we get:
\begin{equation}\label{Inl_R}
{\mathcal I}_{\rm nl} \simeq \frac{C_{\rm eff}}{1+C_{\rm eff}}.
\end{equation}

For typical small bare couplings $g_{1,2}$, one has $C_{\rm eff} \ll 1$ and Eq.~(\ref{Inl_R}) gives $\mathcal{I}_{\rm nl} \simeq C_{\rm eff}$.  On the other hand, the possibility of approaching $C_{\rm eff}, {\mathcal I}_{\rm nl} \sim 1$ is very interesting and in a four-mode system it should be much easier to reach this regime. In the two-mode system it was found that the perturbative treatment compares favorably to a numerical solution of the master equation, even if nonlinear effects are very strong \cite{lemonde2013nonlinear}. Therefore, we expect that the quantitative predictions of our theory can still be valid when $C_{\rm eff} \sim 1$. More precise conditions for the validity of our treatment will be given below.

We have already noted that Fig.~\ref{fig_I_eresonance} indicates an interesting enhancement of the $\mathcal{I}_{\rm nl}$ with detuning. To discuss Eq.~(\ref{R}) in the limit of large $|\Delta|/\omega_m$, it is useful to define the parameter $\varg$:
\begin{equation}\label{Gpm_largeD}
\varg =G_-/ G_{\rm cri},
\end{equation}
which is in the allowed interval $\sqrt{3}/2 \lesssim \varg <1$ when $|\Delta|/\omega_m \gg 1$  (see also Fig.~\ref{fig_I_eresonance}). At large detuning, the other dressed coupling and the polariton frequencies are given respectively by $G_+\simeq \sqrt{4\varg^2 -3}G_{\rm cri}$ and $\omega_{-,\pi} =  \omega_{-,2\pi}/2 \simeq  \omega_m \sqrt{1-\varg^2}$. The polariton decay rates are found as follows:
\begin{align}\label{kappa12}
\kappa_{-,\pi}&\simeq \varg^2 \frac{\omega_m^2}{|\Delta|^2}\kappa  +\frac{\gamma}{\sqrt{1-\varg^2}}, \nonumber \\
\kappa_{-,2\pi}&\simeq (4\varg^2-3)\frac{\omega_m^2}{|\Delta|^2}\kappa +\frac{\gamma}{2\sqrt{1-\varg^2}},
\end{align}
and the asymptotic expressions for $\tilde{g}_e$ and $n_{-,\pi}$ are (see details in Appendix~\ref{appendix1}):
\begin{align}
\tilde{g}^2_e &\simeq  \frac{9(4\varg^2-3)c(\varg)^2}{16(1-\varg^2)^{3/2}} \left(\frac{\omega_m}{|\Delta|}\right)^2 g_1^2, \label{g_asympt} \\
n_{-,\pi}&\simeq \frac{|\Delta|/\omega_m}{4\sqrt{1-\varg^2}}\left(1 +\frac{\gamma}{\kappa}\frac{|\Delta|^2/\omega_m^2}{\varg^2\sqrt{1-\varg^2}}\right)^{-1}, \label{n_asympt} 
\end{align}
where $c(\varg)=\varg^2/(\varg+\sqrt{4\varg^2-3})$ is a numerical factor of order unity. In Eq.~(\ref{n_asympt}) we assumed $T=0$. 

\begin{figure}
\centering
\includegraphics[width=\columnwidth]{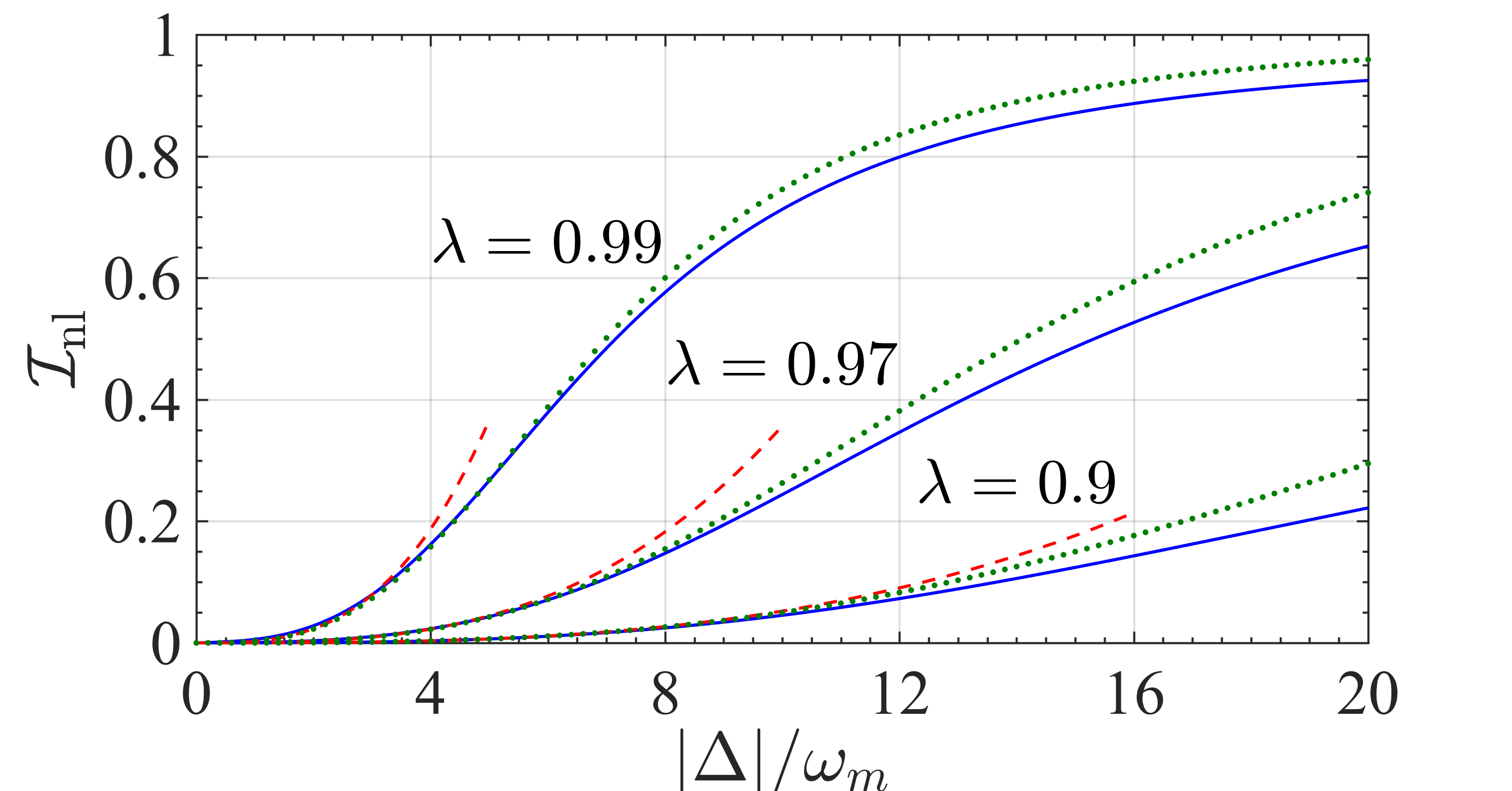}
\caption{(Color online) Dependence of ${\cal I}_{\rm nl}$ on $|\Delta|/\omega_m$ at several values of $\varg=G_{-}/G_{\rm cri} = 0.9, 0.97, 0.99$ (bottom to top). Other parameters are as in the right panel of Fig.~\ref{fig: nonlinearStr_Gp_Gm}. When ${\cal I}_{\rm nl} \ll 1$, the full results (solid) are in agreement with the approximated $C_{\rm eff}$ (dashed, red), given by Eq.~(\ref{R_approx}). We also use Eq.~(\ref{R_approx}) to plot $C_{\rm eff}/(1+C_{\rm eff})$ (dotted, green). Beyond the range of the plot, the agreement will be lost and $\cal{I}_{\rm nl}$ eventually decreases to zero. See Eqs.~(\ref{neglect_gamma}) and (\ref{Delta_opt}), and related discussions. }
\label{Inl_vs_Delta}
\end{figure}

Combining these results and neglecting the (usually small) mechanical damping $\gamma$ we finally reach the important formula:
\begin{equation}\label{R_approx}
C_{\rm eff} \simeq \frac98 \left( \frac{c(\varg)/\varg}{1-\varg^2}\right)^2 \left(\frac{|\Delta|}{\omega_m}\right)^3 \frac{g_1^2}{\kappa^2}.
\end{equation}
In deriving this expression we have also used $1+2n_{-,\pi} \simeq 2n_{-,\pi}$, which is justified at large $|\Delta|/\omega_m$ and $\varg \simeq 1$. 

As shown in Fig.~\ref{Inl_vs_Delta}, Eq.~(\ref{R_approx}) describes well the nonlinear effects. Figure~\ref{Inl_vs_Delta} also demonstrates the potential of observing very large nonlinear effects, and even reach $C_{\rm eff} >1$. Notice that $C_{\rm eff} = 1$ gives $\mathcal{I}_{\rm nl} \simeq 1/2$, i.e., a $50\%$ change in the DOS.

\subsection{Enhancement induced by detuning}

A strong dependence on $(|\Delta|/\omega_m)^3$ appears in Eq.~(\ref{R_approx}) and, by making use of Eqs.~(\ref{kappa12}-\ref{n_asympt}), we can trace its origin. As it turns out, the largest power is contributed by the effective dampings since:
\begin{equation}
\kappa_{-,\pi} \sim  \kappa_{-,2\pi}\propto  \left(\frac{\omega_m}{|\Delta|} \right)^2 \kappa  ,
\end{equation}
giving a factor $(\omega_m/|\Delta|)^4$ in the denominator of Eq.~(\ref{R}). The enhancement from the effective dampings is partially compensated by the decrease of the effective coupling:
\begin{equation}
\tilde{g}_e^2 n_{-,\pi} \propto \left(\frac{\omega_m}{|\Delta|} \right) g_1^2,
\end{equation}
which reflects the quadratic suppression of $\tilde{g}^2_e$ and a linear enhancement of quantum heating, $n_{-,\pi}  \propto |\Delta|/\omega_m$.

The physical origin of these dependences can be understood by noticing that when $|\Delta|/\omega_m \gg 1$ the coupling between optical and mechanical modes is much smaller than the gap $|\Delta|$: from Eq.~(\ref{Gpm_largeD}) we see that $G_\pm \sim G_{\rm cri}=\sqrt{|\Delta|\omega_m}/2  \ll |\Delta|$. Thus, the two low-energy polaritons are mechanical modes with a very weak coupling to the optical modes. This is why the polariton decay rates can be effectively reduced by a larger detuning, which suppresses the mixing with the much strongly damped cavities. On the other hand, a large $|\Delta|$ is also suppressing the effective coupling, as purely mechanical modes do not interact with each other.

It should be stressed that the above discussions neglect the role of mechanical damping $\gamma$, which is only justified when:
\begin{equation}\label{neglect_gamma}
\frac{|\Delta|^2}{\omega_m^2} \ll \left(\frac{\sqrt{1-\varg^2}}{4\varg^2-3}\right) \frac{\kappa}{\gamma}.
\end{equation}
Typically, $\gamma \ll \kappa$ and the condition in Eq.~(\ref{neglect_gamma}) is not very restrictive, except when approaching the instability ($\varg \to 1$). However, it is instructive to check what happens when Eq.~(\ref{neglect_gamma}) is violated: since the dampings saturate to $\kappa_{-,\pi} \sim  \kappa_{-,2\pi}\sim  \gamma$ and the dependence of $C_{\rm eff}$ and $\cal{I}_{\rm nl}$ is dominated by the quickly vanishing effective coupling $\tilde{g}_e^2 n_{-,\pi} \sim (\omega_m/|\Delta|)^3 g_1^2 \kappa /\gamma$, nonlinear effects disappear if $|\Delta|$ is too large. The optimal detuning can be estimated as:
\begin{equation}\label{Delta_opt}
\Delta_{\rm opt} \sim -\omega_m\sqrt{\kappa/\gamma} .
\end{equation} 
The best working point, however, might be dictated by other considerations, e.g., limitations on the largest achievable couplings (since $G_\pm \sim \sqrt{|\Delta|\omega_m}$).

\subsection{Enhancement towards the unstable regime}

Another possible strategy to enhance $\cal{I}_{\rm nl}$, suggested by Eq.~(\ref{R_approx}), is to tune parameters in proximity of the instability. In fact, Eq.~(\ref{R_approx}) shows a divergence in the limit $\varg\to 1$. At variance with the previous discussion, this effect is not due to the polariton linewidths (which are approximately independent of $\varg$) but can be traced down to the divergence in effective coupling and occupation, respectively given by Eqs.~(\ref{g_asympt}) and~(\ref{n_asympt}). 

In principle, this would allow to achieve large nonlinear effects even at fixed moderate detuning. However, the perturbative treatment will eventually break down at the instability. Notice that, to discuss the validity of our theory at $\varg \to 1$, we should take into account not only the larger effective coupling, but also the vanishing polariton energies. Since $\omega_{-,2\pi}=2\omega_{-\pi} \to 0$, the linewidths of the two relevant polaritons can approach their energy separation. We should require $\omega_{-,\pi} \gg \kappa_{-,2\pi},\kappa_{-\pi}$ or, equivalently:
\begin{equation}\label{narrow_linewidth}
\sqrt{1-\varg^2} \gg  {\rm max}\left[\frac{\omega_m \kappa}{|\Delta|^2},\sqrt{\frac{\gamma}{\omega_m}}\right].
\end{equation}
For moderate detuning, the first quantity in the square brackets is usually larger. For example, if we assume $|\Delta|/\omega_m \sim 3$ and a rather typical value $\gamma/\omega_m \sim 10^{-6}$, the inequality $\omega_m \kappa/|\Delta|^2>\sqrt{\gamma/\omega_m}$ implies $\kappa/\omega_m \gtrsim 10^{-2} $, which is usually the case. In this regime we have $\sqrt{1-\varg^2} \gg  \omega_m \kappa/|\Delta|^2 > \sqrt{\gamma/\omega_m}$ and $\varg \simeq 1$, implying that also Eq.~(\ref{neglect_gamma}) is satisfied.

Unless we approach very closely the unstable regime and/or the optical damping $\kappa$ is relatively large, the condition of Eq.~(\ref{narrow_linewidth}) does not pose a restriction to the validity of our theory. For example, Eq.~(\ref{narrow_linewidth}) is well satisfied for the results shown in Fig.~\ref{Inl_vs_Delta} where it gives $1-\varg \gg 10^{-4}$ at $|\Delta|=1.5 \omega_m$ (at larger $|\Delta|$ the condition becomes even less restrictive). 

It is also interesting to discuss the largest nonlinear effects which can be obtained before the perturbative treatment breaks down. By taking into account Eq.~(\ref{narrow_linewidth}), Eq.~(\ref{R_approx}) gives:
\begin{equation}\label{R_max}
C_{\rm eff} \ll \frac{9}{32} \left( \frac{\omega_m}{\kappa}\right)^4 \left(\frac{|\Delta|}{\omega_m}\right)^{11} \frac{g_1^2}{\kappa^2},
\end{equation}
where the right-hand-side can be much larger or much smaller than one, depending on system parameters. For example, for $g_1/\kappa =10^{-3}$, $\kappa/\omega_m = 1/10$, and $|\Delta|/\omega_m \sim 3$ (similar to Fig.~\ref{Inl_vs_Delta}), Eq.~(\ref{R_max}) gives $C_{\rm eff} \ll 500$. On the other hand, a system with $\kappa/\omega_m \sim 1$ (barely in the resolved sideband regime) gives a much more restrictive $C_{\rm eff} \ll 0.05$ keeping $g_1/\kappa $ and $|\Delta|/\omega_m$ unchanged.

In summary, while it is clear that a non-perturbative description becomes necessary close to the instability, Eq.~(\ref{R_max}) indicates that at sufficiently small $\kappa/\omega_m$ and $\omega_m/|\Delta|$ our treatment is valid up to $C_{\rm eff}\gg 1$, i.e., when strong nonlinear effects are well established. Generally speaking, a smaller value of $\kappa/\omega_m$ is advantageous in observing the effects of nonlinearity, and also greatly extends the regime of validity of our theory.

\subsection{Spectral properties and OMIT signal} \label{sec_omit}

The nonlinear effects discussed so far could be observed in OMIT experiments \cite{weis2010optomechanically,safavi2011electromagnetically}, where one of the cavities is probed through a weak laser (in addition to the strong drive). The reflection coefficient for the probe can be computed from $\mathcal{G}^R$ as \cite{lemonde2013nonlinear}:
\begin{equation}
r(\omega_{\rm p})=1-i \kappa_{\rm cp} \mathcal{G}^R ( d_1, d^{\dagger}_1; \omega_{\rm p} ),
\end{equation}
where $\kappa_{\rm cp} < \kappa$ is the contribution to the cavity damping from coupling to the input/output modes. In fact, the OMIT signal is strictly related to the density of states $\rho_d[\omega]$: at small $\kappa_{\rm cp}$, it is easy to show that
\begin{equation}
\left|r(\omega_{\rm p})\right|^2 \simeq 1- 2\pi \kappa_{\rm cp} \rho_d[\omega_{\rm p}].
\end{equation}

\begin{figure}
\centering
\includegraphics[width=\columnwidth]{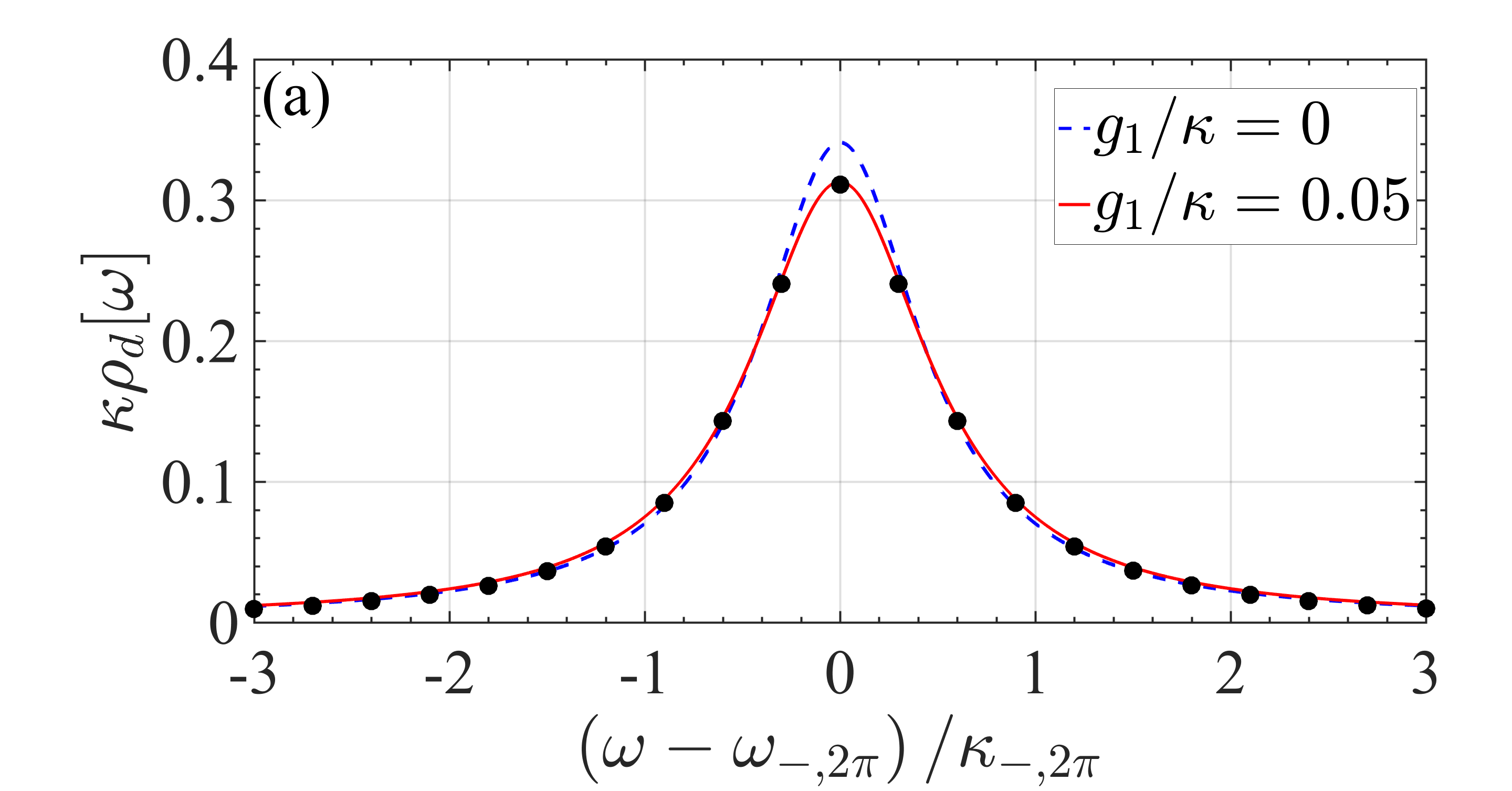}
\includegraphics[width=\columnwidth]{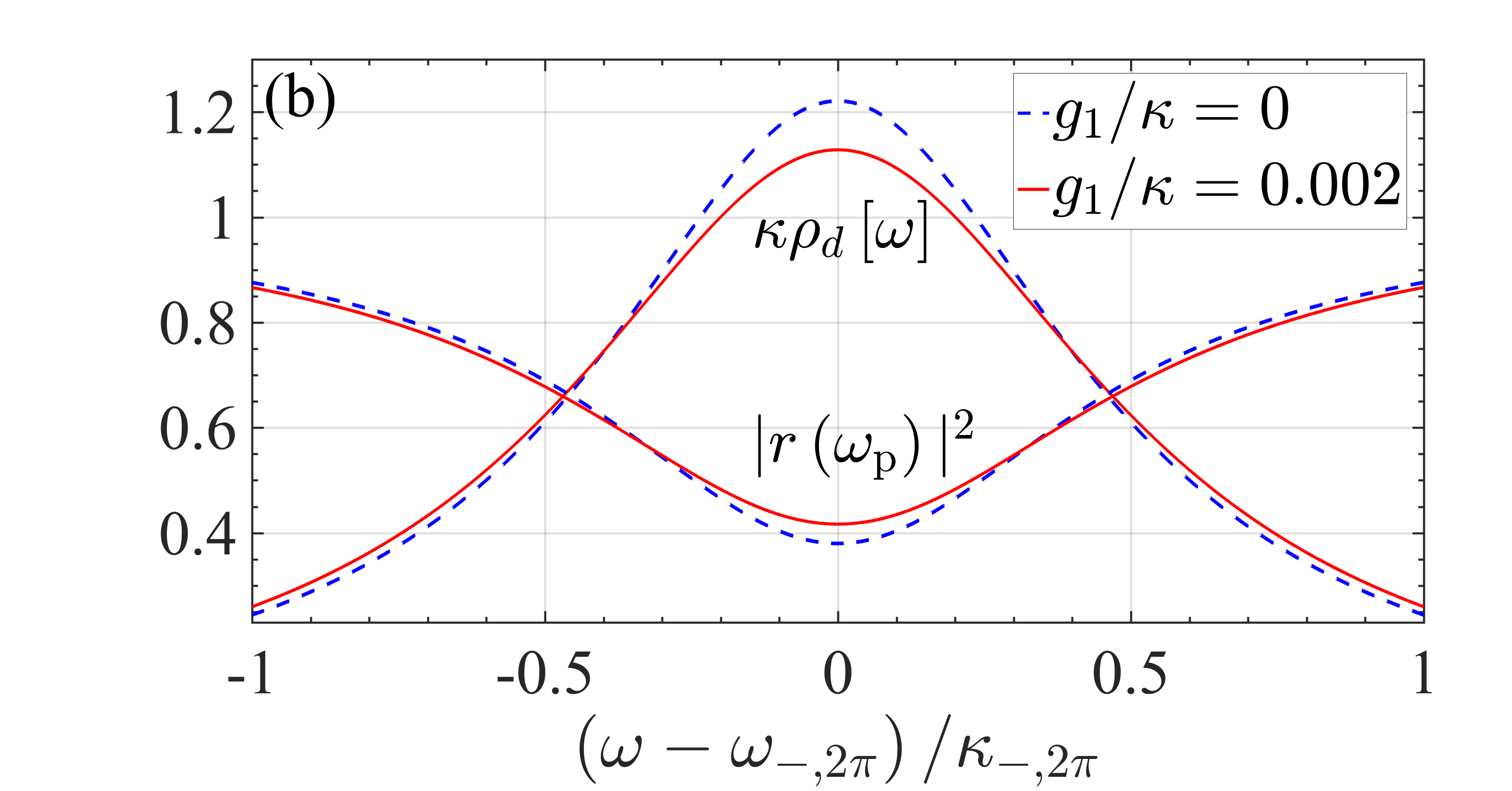}
\caption{ (Color online) (a) A comparison of cavity DOS in the linear (dashed, blue) and nonlinear (solid, red) regimes. The black points are obtained from a numerical solution of the master equation for the polariton modes, including nonlinear interactions. (b) Cavity DOS and OMIT reflection probability. The dashed blue (solid red) curves correspond to the linear (nonlinear) regime. In (a) we used $\Delta =- 1.5 \omega_m$ and $G_{-} =0.52 \omega_m$. In (b) $\Delta =-8 \omega_m$, $G_{-} =1.35 \omega_m$, and $\kappa_{\rm cp}/\kappa =0.1$. Other parameters are as in the right panel of Fig.~\ref{fig: nonlinearStr_Gp_Gm}.}
\label{fig:frequency}
\end{figure}

For parameters satisfying resonant condition (e), we show in Fig.~\ref{fig:frequency} typical examples of the frequency dependence around $\omega_{-,2\pi}$, both for the DOS and the OMIT signal. In panel (a) of Fig.~\ref{fig:frequency} we also provide a comparison of the Keldysh approach with a direct numerical solution of the master equation \cite{johansson2012qutip} (black points). We find good agreement between the two approaches.

At variance with the two-mode system \cite{lemonde2013nonlinear,borkje2013signatures}, the spectrum in Fig.~\ref{fig:frequency} remains qualitatively the same in the presence of nonlinearity. There is a suppression of the peak (dip) in the DOS (reflectivity) of the four-mode chain, but only very large values of the nonlinearity can result in a splitting of the polariton peak (dip). The difference can be understood by considering the resonant structure of the self-energy Eq.~(\ref{sigma:1a}) and the relative size of the effective dampings. Since in our four-mode chain we have $ \kappa_{-,\pi} \geq \kappa_{-,2\pi}$ [from Eq.~(\ref{kappa12})], nonlinear effects occur in a range of frequencies broader than the width $\kappa_{-,2\pi}$ of the main peak (dip).

\begin{figure}
\centering
\includegraphics[width=\columnwidth]{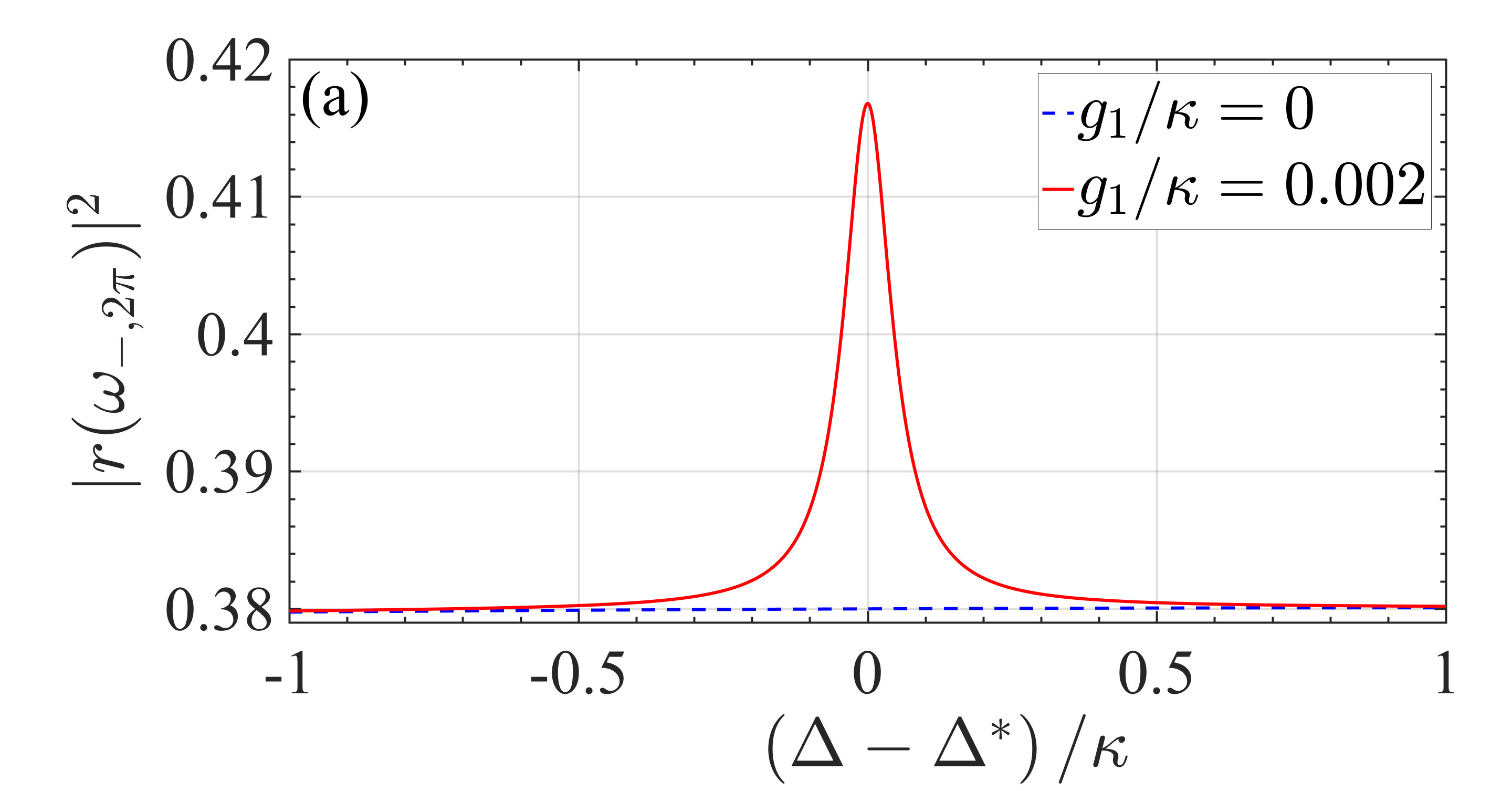}
\includegraphics[width=\columnwidth]{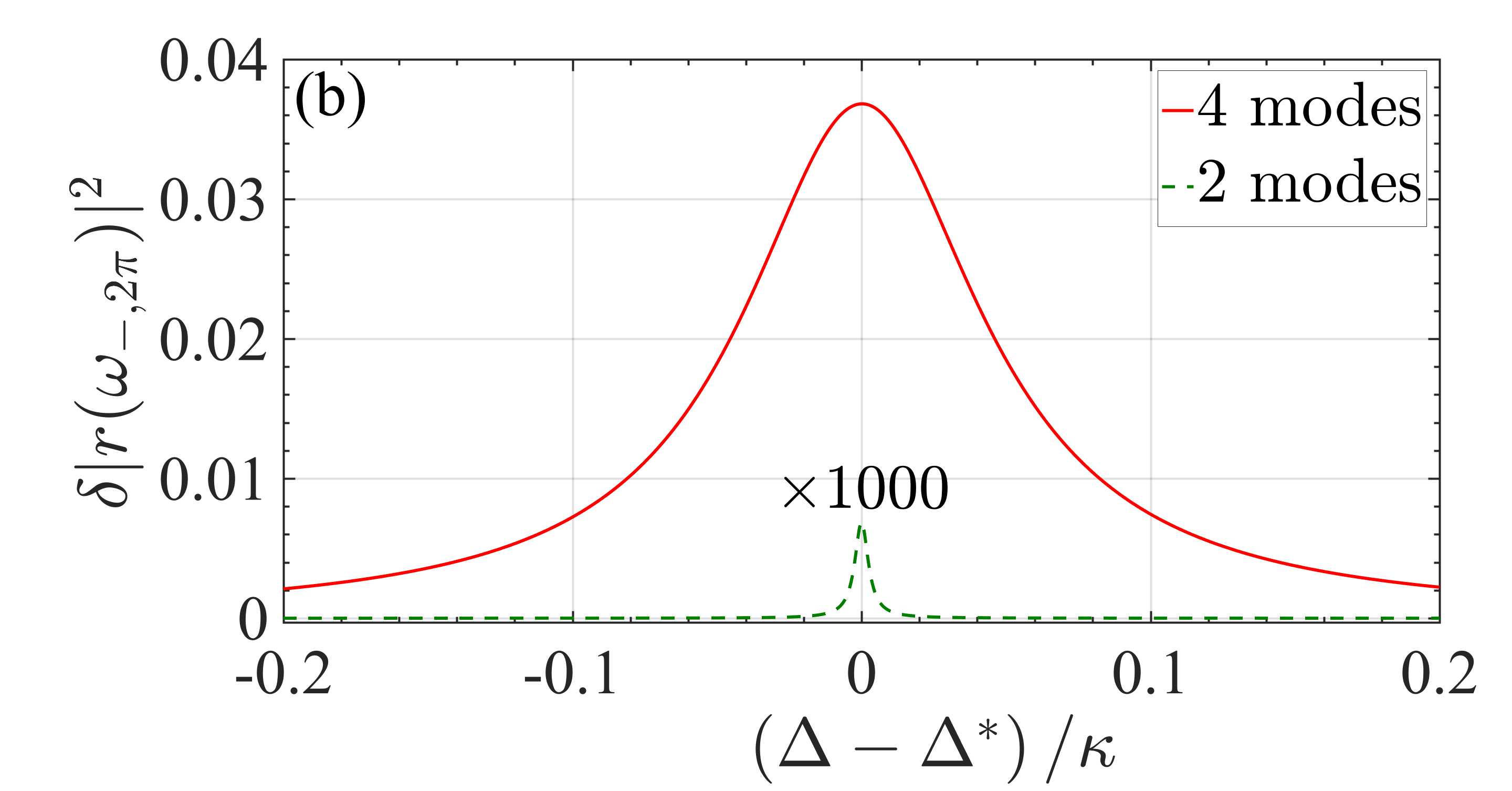}
\caption{ (Color online) Dependence of reflectivity on detuning. Panel (a) shows that nonlinear effects give rise to a pronounced peak in the minimum reflectivity around $\Delta^* = - 8 \omega_m$, when resonance (e) is realized. Other parameters: $\kappa/\omega_m = 0.1$, $\gamma/\kappa = 10^{-4}$, $G_-/\omega_m  = 1.35 $, and $G_+/\omega_m  = 1.137 $. (b): Comparison of the nonlinear signatures of panel (a) to a two-mode optomechanical system with the same values of $\kappa,\gamma$, and $g=g_1$. The detuning of the two-mode system is around $\Delta^* = -1.99\omega_m$ (to maximize nonlinear effects), giving $G/\omega_m= 0.0545$ for the dressed coupling. In (b) we plot $\delta|r(\omega_{-,2\pi})|^2=|r(\omega_{-,2\pi})|^2-|r_0(\omega_{-,2\pi})|^2$, where $r_0$ neglects the nonlinearity. }
\label{detuning_peak}
\end{figure}

For this reason, instead of the spectrum at the resonant condition, it might be easier to probe the sensitive dependence of nonlinear effects on various tunable parameters. For example, in the case of Fig.~\ref{fig: nonlinearStr_Gp_Gm} several sharp peaks will be observed by tuning the system along the dashed line, i.e., by changing the strength of the drive (see also Fig.~\ref{line_cut}). In a similar way, we show in Fig.~\ref{detuning_peak} that sharp features can be observed at fixed $G_\pm$ as a function of detuning. If resonance (e) occurs at detuning $\Delta^*$, we obtain:
\begin{equation}\label{Ceff_lorenzian}
C_{\rm eff} (\Delta)\simeq \frac{\Gamma^2/4}{(\Delta -\Delta^*)^2 + \Gamma^2/4}C_{\rm eff} (\Delta^*),
\end{equation}
where:
\begin{equation}\label{Lambda}
\Gamma = \frac{8\sqrt{1-\varg^2}}{3} \frac{\omega_m}{|\Delta^*|} \kappa.
\end{equation}
We derived Eq.~(\ref{Ceff_lorenzian}) from Eq.~(\ref{R}), by expanding $\omega_{-,2\pi}-2\omega_{-,\pi}$ (appearing in the self-energy) at small values of $(\Delta-\Delta_*)$. We find that this approximation is in good agreement with the full expressions and, in particular, Fig.~\ref{detuning_peak} is described very well by Lorenzian peaks of width given by Eq.~(\ref{Lambda}). 

In Fig.~\ref{detuning_peak}(b) we also provide an example comparing the two- and four-mode systems, by assuming the same values of $\kappa, \gamma$, and nonlinear coupling strength in the two cases. For this particular choice of parameters, the enhancement of nonlinear features is of about four orders of magnitude.

\section{Comparison to the two-mode system} \label{sec_2mode}

We have seen that the enhancement induced by process (e) is analogous to what happens in a two-mode system \cite{lemonde2013nonlinear,borkje2013signatures}. However, the last example of the previous section shows that there can be a large quantitative difference between the two cases. Here we would like to clarify the origin of the enhancement.

Our $(-,2\pi) $ and $(-,\pi) $ modes play the same role of the upper ($+$) and lower ($-$) polaritons of the two-mode case, respectively, and our Eq.~(\ref{R}) corresponds to: 
\begin{equation}
C_{\rm eff} = \frac{4\tilde{g}^2(1+2n_-)}{\kappa_- \kappa_+}. ~~~ {\rm (two~modes)}
\end{equation}
The precise definition of $\tilde{g}$, $n_-$, and $\kappa_\pm$ can be found in Ref.~\cite{lemonde2013nonlinear}. It can also be inferred by our expressions of Sec.~\ref{Sec: linear}, usually by dropping the subscript $k$.

In the two-mode system, the non-linear effects can be maximized by tuning $\Delta$ but the range of values is restricted to the interval $(-2 \omega_m, - \omega_m/2)$. At the optimal point, which is slightly above $\Delta = -2\omega_m $, we have:
\begin{equation}\label{gk_2mode}
\tilde{g}^2 \approx  \left(G/\omega_m\right)^2 g^2 , \quad
\kappa_- \simeq \frac{8}{9} \left(G/\omega_m\right)^2 \kappa,
\end{equation}
where the factor $G/\omega_m$ (with $G\to 0$) reflects the small mixing angle $\theta$ between the mechanical and optical modes. Similarly to our case, at the optimal point the `optical' and `phononic' polaritons are nearly decoupled. In fact, $G/\omega_m$ is the analog of our $\omega_m/|\Delta| $, since Eq.~(\ref{theta_k}) gives $\theta \sim G/\omega_m $ for the two-mode system and $\theta_k \sim \omega_m/|\Delta| $ for our four-mode chain. 

The main difference between the two cases is that in the two-mode system there can be only one polariton which is mostly phononic, while in the four-mode case both modes involved in the scattering process can be phononic. For this reason, $\kappa_+ \simeq \kappa$ in the two-mode system, therefore the damping of the $+$ polariton does not contribute any enhancement factor to $C_{\rm eff}$. Furthermore, there is no enhancement of $n_-$, which can be easily seen from Eq.~(\ref{occuptions2}): due to the restricted range of $|\Delta|$, we have $|\Delta|\sim \omega_- \sim \omega_m$, and the prefactor of Eq.~(\ref{occuptions2}) is a simple constant. The precise result is $n_- \simeq 1/8$, giving: 
\begin{equation}\label{C2modes}
C_{\rm eff} \simeq \frac{45}{8}  \frac{g^2}{\kappa^2}, ~~~ {\rm (two~modes)}
\end{equation}
which can be compared directly to Eq.~(\ref{R_approx}).

 In the two-mode system, the large difference in damping rates $\kappa_- \ll \kappa_+ $ has the advantage of producing very narrow features in the spectrum. As we discussed, this is not the case of the four-mode chain when $ \kappa_{-,\pi} \geq \kappa_{-,2\pi}$. On the other hand, the relative size of the nonlinear effects can be very different. In Eq.~(\ref{C2modes}), the small factors from Eq.~(\ref{gk_2mode}) have canceled out, and there is no analog of our $(|\Delta|/\omega_m)^3$ dependence. 

Also $(1-\varg^2)^{-2}$ is absent from Eq.~(\ref{C2modes}), since in the two-mode system there is a restricted freedom in the choice of parameters: its resonant condition is indicated by the dot-dashed curve of Fig.~\ref{fig: regime_resonant} (top right panel), which is always far from $G_{\rm cri}$. 

Thus, the added flexibility of the four-mode system has the potential of enhancing a small $g^2/\kappa^2 $ (e.g., $g^2/\kappa^2\sim 10^{-6}$) by large prefactors. As shown in Fig.~\ref{Inl_vs_Delta}, this would bring $C_{\rm eff}$ and $\mathcal{I}_{\rm nl}$ to values which are much more accessible.

\section{Specific implementation}\label{sec: Experimental realization}

In this section we demonstrate an implementation of the model. Note that for $N=2$ each mechanical mode interacts with both optical modes, thus the actual geometry is not restricted by the ring structure. One has only to make sure that the four optomechanical couplings are equal in pairs like in Fig.~\ref{fig: the system}.

\begin{figure}
\centering
\includegraphics[width=\columnwidth]{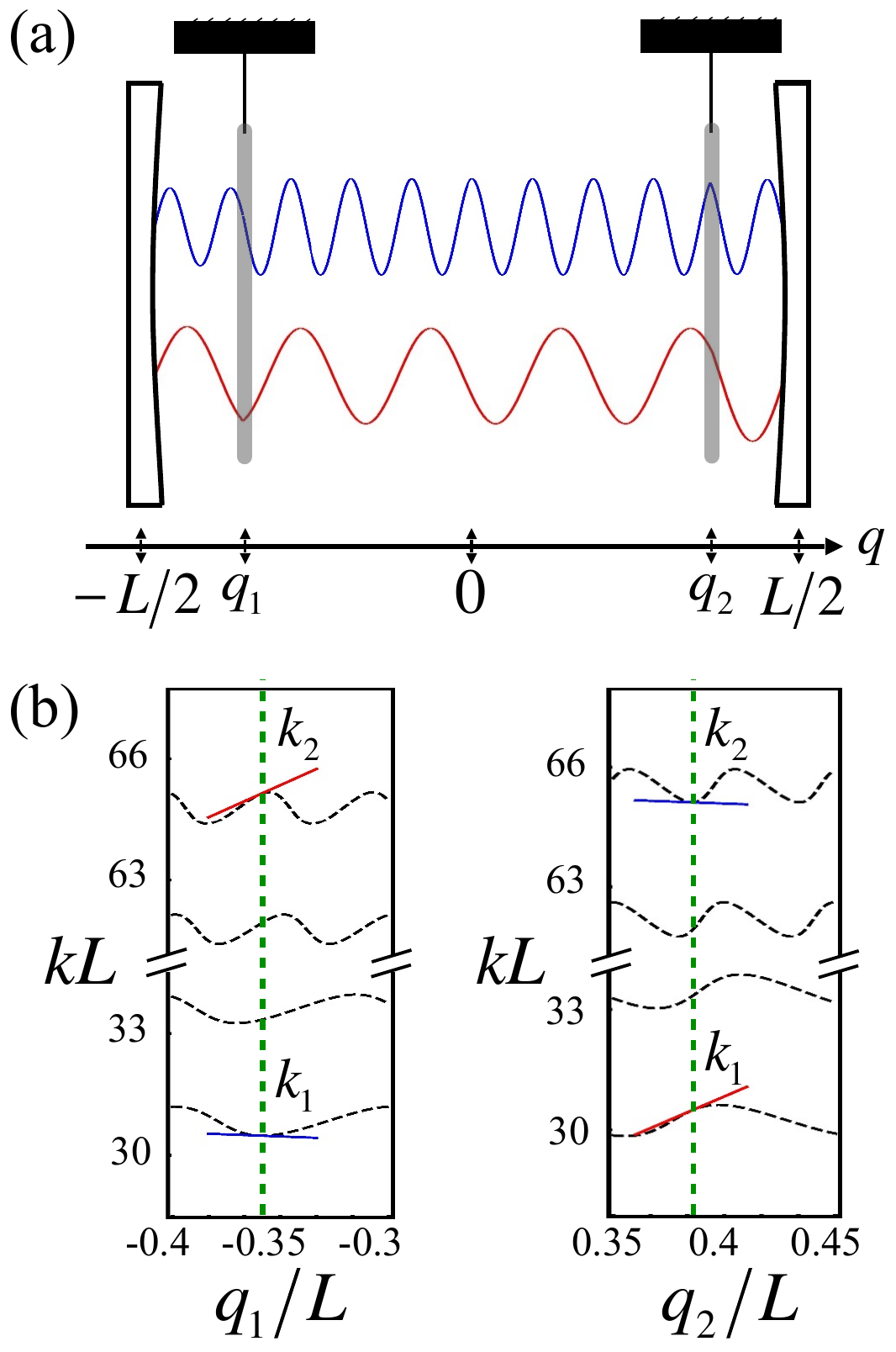}
\caption{(Color online) Example of Fabry-Perot cavity realizing our model. Panel (a) illustrates the position of the mirrors and the spatial dependence of the two relevant optical modes. In (b) we show the optical spectrum as a function of position of the first (left panel) and second (right panel) mirrors. The (green) vertical dashed lines mark the equilibrium positions of the two mirrors. The $k_1$ ($k_2$) mode is at the bottom (top) of the shown spectrum. Parameters are chosen to satisfy $\partial k_2/\partial q_1 = \partial k_1/\partial q_2$ (red positive slopes) and $\partial k_1/\partial q_1 = \partial k_2/\partial q_2$ (blue negative slopes). We used ${\cal T}=0.85$.}
\label{FP_cavity}
\end{figure}

The specific example we consider here is a Fabry-Perot cavity with two identical membranes placed at positions $q_{1,2}$ \cite{bhattacharya2008multiple,hartmann2008steady}. As indicated in Fig.~\ref{FP_cavity}(a), the cavity length is $L$ and we take $q=0$ at the center of the cavity. To obtain the optical spectrum, we model the membranes as dielectric `bumps' \cite{spencer1972theory} with transmission coefficient $\cal T$. After imposing appropriate boundary conditions at the mirrors and membranes, the linear system for the optical amplitudes gives the secular equation:
\begin{align}\label{eq:spectrum equation}
&\sin (k_i L+2\varphi)+\sin[k_i L+2k_i(q_{1}-q_{2})] \sin^{2}\varphi\nonumber\\
&-2\sin\varphi\cos[k_i (q_{1}-q_{2})-\varphi]\cos [k_i (q_{1}+q_{2})]=0, 
\end{align}
where $k_i$ is the wavevector for the $i$-th optical mode and $\varphi= \arccos (\sqrt{{\cal T}})$.

We focus on two of the optical modes, with wavevectors $k_{1,2}$, and compute the bare optomechanical couplings $g_{ij}$, which describe the coupling between the $i$-th optical mode and the $j$-th mechanical mode. They are given by  $g_{ij}= c (\partial k_i/\partial q_{j}) X_{\rm ZPF}$, where $c$ is the speed of light and $X_{\rm ZPF}$ is the zero-point fluctuation of the membranes \citep{aspelmeyer2014cavity}. Using  Eq.~(\ref{eq:spectrum equation}) to rewrite the partial derivative yields:
\begin{eqnarray}\label{g_ij}
g_{ij}= c \frac{B_j(k_i)}{A(k_i)} X_{\rm ZPF},
\end{eqnarray}
where
\begin{align}
A(k_{i})  = & L\cos (k_{i}L+2\varphi)+\tilde{L} \cos (k_{i}\tilde{L})\sin^{2}\varphi \nonumber \\
&+2 \left[q_{1}\sin(2k_{i}q_{1}-\varphi) +q_{2}\sin(2k_{i}q_{2}+\varphi)\right]\sin\varphi,\nonumber \\
B_{1}(k_{i})  = & -2k_{i} \left[\sin(2k_{i}q_{1}-\varphi)+\cos(k_{i}\tilde{L})\sin\varphi\right]\sin\varphi, \nonumber\\
B_{2}(k_{i})  = & -2k_{i}\left[\sin(2k_{i}q_{2}+\varphi)-\cos(k_{i}\tilde{L})\sin\varphi\right]\sin\varphi.
\end{align}
Here, to simplify the notation, we have defined $\tilde{L}=L+2(q_{1}-q_{2})$. Finally, the ring configuration discussed here requires the following relations:
\begin{align}
& g_{12}=g_{21}\equiv g_{1},\\
& g_{11}=g_{22} \equiv g_{2},
\end{align}
or, equivalently:
\begin{eqnarray}
B_{1}(k_{1})A(k_{2}) &=& B_{2}(k_{2})A(k_{1}), \label{g1eq} \\
B_{1}(k_{2})A(k_{1}) &=& B_{2}(k_{1})A(k_{2}).  \label{g2eq}
\end{eqnarray}

By solving simultaneously Eqs.~(\ref{eq:spectrum equation}),~(\ref{g1eq}),~and~(\ref{g2eq}), we can find suitable configurations reproducing Eq.~(\ref{H}). An example is shown in Fig.~\ref{FP_cavity}. Although the wavevectors $k_1$ and $k_2$ of the optical modes are different, we can still realize Eq.~(\ref{H}) by applying driving lasers with identical detunings.

\begin{figure}
\centering
\includegraphics[width=\columnwidth]{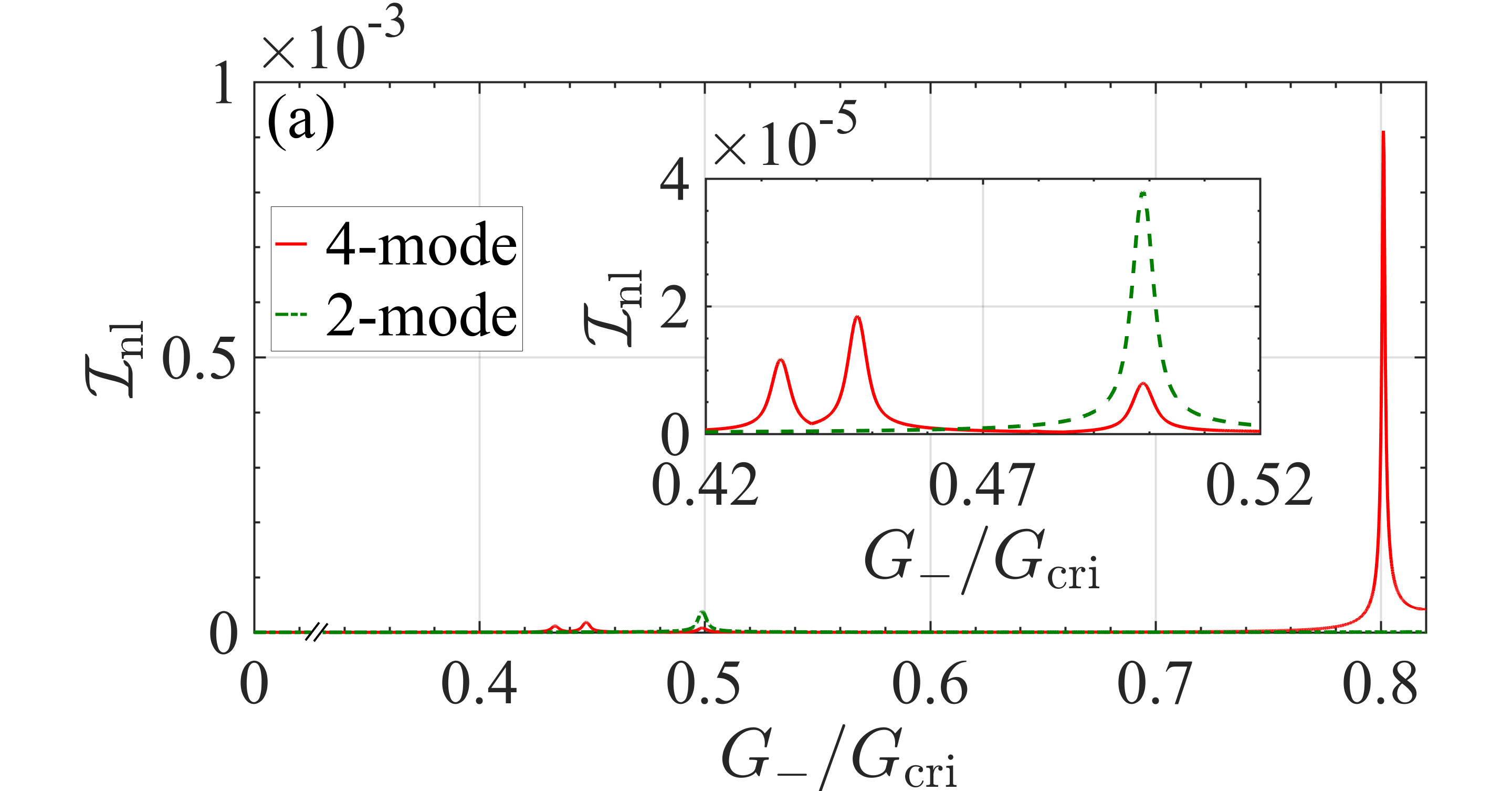}
\includegraphics[width=\columnwidth]{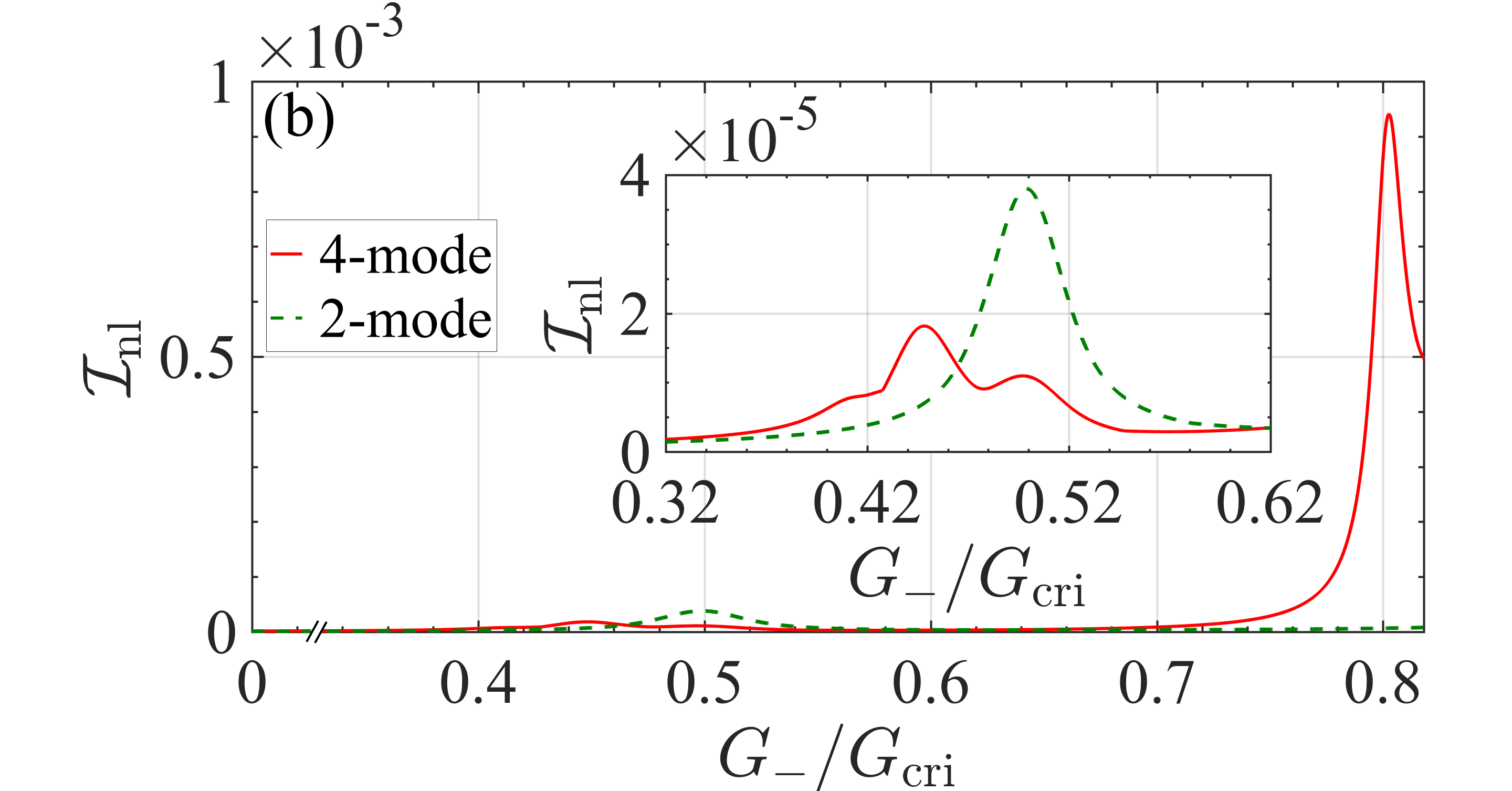}
\caption{(Color online) Plot of $\mathcal{I}_{\rm nl}$ along the line cuts highlighted in Fig.~\ref{fig: nonlinearStr_Gp_Gm}. (a) refers to the left panel of Fig.~\ref{fig: nonlinearStr_Gp_Gm} and (b) to the right panel. Solid red lines are for $G_{-}=1.187G_{+}$ (from the setup of Fig.~\ref{FP_cavity}), while the dashed green lines represent the two-mode system ($G_{-}=G_{+}$). The nonlinear effects in the four-mode system are much stronger, due to the large peak from resonance (e). The insets are zoom-ins of the weaker features induced by resonances (a), (c), and (b).
}
\label{line_cut}
\end{figure}

If we keep the geometry unchanged, the ratio $g_1/g_2$ will be fixed and determines $G_-/G_+$ through Eq.~(\ref{ratio_Gg}). In Fig.~\ref{FP_cavity}, we obtain $G_{-}/G_{+} \approx 1.187$, which corresponds to the (white) dashed line of Fig.~\ref{fig: nonlinearStr_Gp_Gm}. We plot in Fig.~\ref{line_cut} the strength of nonlinear effects ${\cal I}_{\rm nl}$ along this particular line cut, where the value of $G_\pm$ can be changed through the incident power of the driving lasers. There are four peaks in the nonlinear signal ${\cal I}_{\rm nl}$ which, as indicated in Fig.~\ref{fig: nonlinearStr_Gp_Gm}, occur at the resonant conditions (a), (c), (b), and (e). The most prominent peak at $G_- \simeq 0.8 G_{\rm cri}$ is due to resonance (e). In terms of the original couplings, it occurs at $G_1 \simeq 0.54 \omega_m$ and $G_2 \simeq -0.05 \omega_m$. 

As a comparison, we show in Fig.~\ref{line_cut} the dependence of ${\cal I}_{\rm nl}$ assuming $G_-= G_+$ (i.e., along the yellow dot-dashed line of Fig.~\ref{fig: nonlinearStr_Gp_Gm}). As discussed, at this point the system reduces to two non-interacting optomechanical cells thus there is a single peak whose size is comparable to the previous (a), (c), (b) resonances and much smaller than the (e) peak. In this specific example (with moderate detuning $\Delta = -1.5 \omega_m$) the enhancement factor is $\sim 20$, but the available freedom of adjusting $\Delta$ can be used to further facilitate the detection of nonlinearities in the four-mode chain.

\section{Conclusion}\label{Sec: summary}

We have investigated the enhancement of nonlinear effects in optomechanical rings and, in particular,
we computed the cavity DOS and OMIT reflection probability of a four-mode system using Keldysh Green's function approach.  We have found pronounced nonlinear effects when $\omega_{-,2\pi}=2 \omega_{-,\pi}$, i.e., when the dominant nonlinear process is a resonant scattering between two low-dissipation phononic polaritons. 

In order to obtain sharp nonlinear signatures, we propose to tune the detuning or intensity of the driving lasers across the resonant condition. We have shown that at larger detuning and many-photon coupling the nonlinear effects are strongly enhanced with respect to the two-mode case. Therefore, the nonlinear features can be large even at weak bare couplings. 

We have also provided a specific implementation of our model. However, since the origin of the enhancement is not directly related to the ring geometry, we expect that other setups with at least two mechanical modes can exploit effectively the same mechanism. 

Finally, our multi-mode optomechanical system is compatible with more direct approaches developed for two-mode systems. For example, larger bare couplings could be achieved with multi-scatterer mechanical elements~\cite{Xuereb2012,Chesi2015,LiXuerebVitali}, and a strongly reduced optical damping was recently realized through a feedback--controlled drive~\cite{Rossi2017,Rossi2018}. Implementing these or other strategies in a ring geometry would further enhance the nonlinear signatures discussed here.

\color{black}

\begin{acknowledgments}
LJJ is supported by the China Postdoctoral Science Foundation
(Grant No. 2016M600908). YDW acknowledges support from NSFC (Grants No.~11574330 and No.~11434011) and MOST (Grant No.~2017FA0304500). SC acknowledges support from NSFC (Grants No.~11574025, No.~U1530401, and No.~11750110428). It is our pleasure to thank helpful discussions with A. A. Clerk, M.-A. Lemonde, P. Rabl, and D. Vitali.
\end{acknowledgments}

\appendix

\section{Effective bare couplings} \label{appendix1}

We present here the expressions of the effective bare couplings, which are too cumbersome to include in the main text. By considering the five possible resonant processes of Fig.~\ref{fig: regime_resonant}, Eqs.~(\ref{Hamiltonian: nonlinear}) and (\ref{linear_transf}) give:
\begin{widetext}
\begin{align}
&\tilde{g}_{a} = \frac{g_{+}}{\sqrt{2}} ( V''_{12} + V''_{14}) V'_{21} V'_{23}
+\frac{g_{-}}{\sqrt{2}} (V'_{11}+ V'_{13} ) (V''_{22} V'_{21} + V''_{24} V'_{23}),\\
&\tilde{g}_{b} = \frac{g_{+}}{\sqrt{2}}  ( V''_{12} + V''_{14}) V''_{21} V''_{23}
+\frac{g_{-}}{\sqrt{2}}(V''_{11} + V''_{13})(V''_{22} V''_{21} + V''_{24} V''_{23} ), \\
&
\tilde{g}_{c} =\frac{g_{+}}{\sqrt{2}}  (V''_{11} + V''_{13})(V'_{22} V'_{21} + V'_{24} V'_{23} )
+\frac{g_{-}}{\sqrt{2}}\left[ (V'_{11} + V'_{13}) (V'_{22} V''_{21} + V'_{24} V''_{23}) 
+  ( V'_{12} + V'_{14})(V''_{21} V'_{23} + V'_{21} V''_{23} ) \right],\\
&\tilde{g}_{d} = \frac{g_{+}}{\sqrt{2}}  ( V''_{12} + V''_{14})(V'_{22} V'_{21} + V'_{24} V'_{23} )
+  \frac{g_{-}}{\sqrt{2}} \left[  ( V'_{11} + V'_{13})(V'_{22} V''_{22} + V'_{24} V''_{24})
+  (V'_{12} + V'_{14} )(V''_{22} V'_{23} + V'_{21} V''_{24} ) \right],\\
&\tilde{g}_{e} =  \frac{g_{+}}{\sqrt{2}} ( V''_{11} + V''_{13}) V'_{21} V'_{23}+  \frac{g_{-}}{\sqrt{2}} ( V'_{11} + V'_{13})(V''_{21} V'_{21} + V''_{23} V'_{23}) ,  \label{geff_B_2pi}
\end{align}
\end{widetext}
where we defined $g_{\pm}=g_1 \pm g_2$ and introduced shorthand notations for the matrix elements: $V'_{ij}\equiv V_{ij}(\pi) $ and $V''_{ij}\equiv V_{ij}(2\pi) $.

As discussed at length in the main text, process (e) plays a special role and for this case we give the explicit form obtained using Eqs.~(\ref{Vmatrix1}-\ref{Vmatrix4}) for the $V_{ij}(k)$:
\begin{align} \label{geff_e}
\tilde{g}_{e} = & \frac{1}{8} \sqrt{\frac{2\omega_m}{\omega_{-,2\pi}}} \left[
 g_{+} \sin^2 \theta_{\pi} \cos\theta_{2\pi} \left(\frac{|\Delta|}{\omega_{-,\pi}} - \frac{\omega_{-,\pi}}{|\Delta|} \right)
 \right.  \nonumber \\
& \left. + g_{-} \sin 2\theta_{\pi}\sin\theta_{2\pi}
\left(\frac{|\Delta|}{\omega_{-,\pi}} + \frac{\omega_{-,2\pi}}{|\Delta|} \right) 
\right].
\end{align} 
The above Eq.~(\ref{geff_e}) is useful to discuss in a more transparent way the limit of large $|\Delta|/\omega_m$. Several asymptotic behaviors were already given in the main text. In particular, $\omega_{-,\pi} =  \omega_{-,2\pi}/2\simeq  \omega_m \sqrt{1-\varg^2}$. Furthermore, by using $g_1/g_2=G_1/G_2$ and the dependence of $G_\pm$ given in and below Eq.~(\ref{Gpm_largeD}), we find:
\begin{equation}
\frac{g_-}{g_1} \simeq  \frac{2\varg}{\varg+\sqrt{4\varg^2 -3}} , \qquad \frac{g_+}{g_1} \simeq  \frac{2\sqrt{4\varg^2 -3}}{\varg+\sqrt{4\varg^2 -3}}.
\end{equation} 
Finally, the mixing angles can be approximated as:
\begin{align}
\theta_\pi \simeq \varg \frac{\omega_m}{|\Delta|},  \qquad 
 \theta_{2\pi}  \simeq \sqrt{4\varg^2-3} \frac{\omega_m}{|\Delta|},
\end{align}
while $\cos \theta_{2\pi} \simeq 1$. From these relations, it is not difficult to see that Eq.~(\ref{geff_e}) yields Eq.~(\ref{g_asympt}) of the main text.

\section{Formulas for resonances (a-d)}\label{appendix2}

Resonance (a) is realized when:
\begin{equation}\label{Gplus1}
G_+^{(a)}=2\sqrt{\frac{(\omega^2_{-,\pi}-\omega^2_m/4)(\omega^2_{-,\pi}-\Delta^2/4)}{|\Delta|\omega_m}}.
\end{equation}
This condition is identical to $G_+^{(e)}$ [see Eq.~(\ref{Gplus_e})] but must be enforced in a different region, i.e., the (blue) allowed region of Fig.~\ref{fig: regime_resonant}(a). The upper edge is:
\begin{equation}\label{upper_edge_a}
G_{-, {\rm max}}^{(a)}= 
\left\{\begin{array}{cl}
\sqrt{\frac{3|\Delta|}{16\omega_m}\left(\omega_m^2-\frac{\Delta^2}{4}\right)},~& {-2\omega_m \leq \Delta<-\omega_m},\\
\sqrt{\frac{3\omega_m}{16|\Delta|}\left(\Delta^2-\frac{\omega_m^2}{4}\right)},~& {-\omega_m< \Delta \leq -\frac{\omega_m}{2}},
\end{array}
\right.
\end{equation}
and the lower edge is:
\begin{equation}\label{Glower1}
G_{-, {\rm min}}^{(a)}= 
\frac18 \sqrt{\Theta\left(\frac{10\Delta^2\omega_m^2-3\Delta^4-3\omega_m^4}{|\Delta|\omega_m}\right)},
\end{equation}
where the argument of the theta function is nonzero only when $-\sqrt{3}\omega_m \leq \Delta < -\omega_m/\sqrt{3}$. 

Resonance (b), is realized when:
\begin{equation} \label{Gplus2}
G_{+}^{(b)} =\frac{1}{10}\sqrt{\frac{17 \Delta^2 \omega^2_m -4 (\Delta^4 + \omega^4_m)}{\vert\Delta\vert \omega_m}},
\end{equation}
and the (cyan) allowed region of Fig.~\ref{fig: regime_resonant}(b) is given by the conditions $-\omega_m<\Delta < -\omega_m/2$ and $0< G_- < G_{\rm cri}$. Interestingly, Eq.~(\ref{Gplus2}) is independent of $G_-$.

Resonance (c) is realized by:
\begin{equation} \label{Gplus3}
G_{+}^{(c)} =\sqrt{\frac{5|\Delta|\omega_m+16G_-^2}{4}-\frac{\Delta^2+\omega_m^2}{2}\sqrt{1-\frac{4G_-^2}{|\Delta|\omega_m}}}.
\end{equation}
The (green) allowed region of Fig.~\ref{fig: regime_resonant}(c) has an upper edge given by a rescaling of Eq.~(\ref{upper_edge_a}): $G_{-,{\rm max}}^{(c)}=2G_{-,{\rm max}}^{(a)}/\sqrt{3}$.

Resonance (d) is realized by a condition identical to case (c), see Eq.~(\ref{Gplus3}), but a different allowed region shown in black in Fig.~\ref{fig: regime_resonant}(d). The upper edge is simply $G_{-,{\rm max}}^{(d)}=G_{\rm cri}$ and the lower edge is given by a rescaling of Eq.~(\ref{Glower_e}): $G_{-, {\rm min}}^{(d)}=2G_{-, {\rm min}}^{(e)}/\sqrt{3}$.

\bibliography{refs}

\end{document}